\documentclass[11pt,twoside]{article}
\RequirePackage[colorlinks,citecolor=blue,urlcolor=blue,
allcolors=blue]{hyperref}
\usepackage[letterpaper,top=0.75in,left=1in,textwidth=6.5in,
textheight=9.25in]{geometry}
\usepackage{authblk}
\usepackage{graphicx}
\usepackage{epstopdf, epsfig}
\usepackage{amsmath}
\usepackage{amsfonts}
\usepackage{color}
\usepackage{bm}
\usepackage[font={small}]{caption}

\graphicspath{{./Figs/}}

\newcommand{\fbar}{\mathcal{F}}
\newcommand{\fstar}{\mathcal{F}^\star}

\newcommand{\Lf}{\ell}
\newcommand{\Xg}{\bar{\mathbf{X}}}

\begin{document}

\title{\vspace{-40pt} Dynamics and fragmentation of small\\ inextensible
  fibers in turbulence}

\author[$^1$]{Sof\'{\i}a Allende}
\author[$^2$]{Christophe Henry}
\author[$^1$]{J\'er\'emie Bec}
\affil[$^1$]{\small MINES ParisTech, PSL Research
  University, CNRS, CEMEF, Sophia-Antipolis, France}
\affil[$^2$]{\small Universit\'e C\^ote d'Azur, INRIA, Team TOSCA,
  Sophia-Antipolis, France}

\maketitle

\begin{abstract}
  \noindent The fragmentation of small, brittle, flexible,
  inextensible fibers is investigated in a fully-developed,
  homogeneous, isotropic turbulent flow. Such small fibers spend most
  of their time fully stretched and their dynamics follows that of
  stiff rods. They can then break through tensile failure,
  \textit{i.e.}\/ when the tension is higher than a given threshold.
  Fibers bend when experiencing a strong compression. During these
  rare and intermittent buckling events, they can break under flexural
  failure, \textit{i.e.}\/ when the curvature exceeds a
  threshold. Fine-scale massive simulations of both the fluid flow and
  the fiber dynamics are performed to provide statistics on these two
  fragmentation processes. This gives ingredients for the development
  of accurate macroscopic models, namely the fragmentation rate and
  daughter-size distributions, which can be used to predict the time
  evolution of the fiber size distribution.  Evidence is provided for
  the generic nature of turbulent fragmentation and of the resulting
  population dynamics. It is indeed shown that the statistics of
  breakup is fully determined by the probability distribution of
  Lagrangian fluid velocity gradients. This approach singles out that
  the only relevant dimensionless parameter is a local flexibility
  which balances flow stretching to the fiber elastic forces.
\end{abstract}

\section{Introduction}
\label{sec:intro}

The fragmentation process, which consists in breaking a body in
different pieces, is very common and relevant to a wide range of
phenomena in science and technology~\cite{beysens1995fragmentation,
  hufner1986fragmentation}. Natural examples occurring at different
length scales are numerous, from breakup in DNA
chains~\cite{nagata2000apoptotic,yan2006unique} to meteors in
space~\cite{keil1994catastrophic}. Moreover, a precise understanding
of the material properties involved in the breaking process proves
fundamental in several industrial applications, as in
combustion~\cite{seames2003initial} or in wastewater
treatment~\cite{verawaty2013breakage}. Traditionally, fragmentation is
modelled from a macroscopic point of view using statistical approaches
(see, \textit{e.g.}, \cite{horn1984midpoint, aastrom2006statistical,
  griffith1943theory, babler2012breakup}) to predict the time
evolution of fragment size distributions from empirical
observations. It relies on population balance models, which are based
on a set of PDEs giving the mean-field behaviour of a population of
objects. Such models require information on two quantities: the
fragmentation rate (\textit{i.e.}\/ the frequency of the breakup
events) and the daughter size distribution (\textit{i.e.}\/ the size
of all fragments generated by a breakup event). New models for these
two quantities are needed to account for the effects of fluctuations
and fine physical phenomena occurring during breakup.

We focus here on the fragmentation of brittle elongated particles with
a constant length, which will be called inextensible fibers in the
following. From the point of view of material sciences, a brittle
material breaks under the action of an external force with little
elastic deformation and without plastic deformation
\cite{rosler2007mechanical}.  In essence, breakup at the molecular
level occurs when the local stress overcomes the internal cohesion
between molecules. In the case of brittle fragmentation, the fracture
induced by this rupture of equilibrium is assumed to propagate
instantaneously at the material level leading to the fragmentation of
the whole object
\cite{aastrom2006statistical,vandenberghe2013geometry}. At the scale
of the fiber, this fragmentation can occur due to three different
actions: tensile failure occurs when the external force acts to
stretch the fiber along its main axis; flexural failure happens when
an external torque induces a flexion perpendicular to the fiber main
axis; torsional failure occurs when the fiber is twisted by an applied
torque. Specific fragmentation thresholds can be defined for each of
these situations: a stretch force in tensile failure, a bending angle
in flexural failure and a twisting angle in torsional failure.

This study addresses more specifically the case of inextensible fibers
immersed in a fluid. Such situations are found in a number of
applications. For example, in the paper industry, cellulose fibers
have been investigated in~\cite{lundell2011fluid}. In a biological
context, fiber dynamics have been used to model diatom phytoplankton
colonies in the ocean~\cite{ardekani2017sedimentation}, organic
matter at fresh water intakes~\cite{santoso2018analysis}.  Besides,
we consider the case of fibers that are smaller compared to the
smallest fluid scale (the Kolmogorov scale in turbulent flows).  In
that case, the dynamics of a fiber is determined by the action of
three forces: bending elasticity, viscous drag and internal tension.
The fiber will mostly experience tensile and flexural failures due to
its stretching or compression by the flow, while torsional failure is
negligible due to the fluid flow linearity at such small
scales. Bending elasticity and viscous drag act together to stretch
the fiber, making it akin to a stiff rod. Tensile failure then occurs
when the local tension reaches values above a threshold. However, when
fibers change their directions and experience strong-enough
compression, their configurations can become
buckled~\cite{becker2001instability}: this is known as the buckling
instability. The instability has been well documented for simple
steady shear flows~\cite{lindner2015elastic}, in which there exists a
critical value of the flexibility above which buckling
occurs. Clearly, flexural failure can only occur when the fiber
buckles and the curvature overcomes a threshold.

The problem of turbulent fragmentation has been essentially addressed
for droplets~\cite{biferale2014deformation,ray2018droplets}, fractal
flocs~\cite{kobayashi1999breakup}, and microscopic
polymers~\cite{vanapalli2006universal,pereira2012polymer}. The breakup
of macroscopic fibers has been essentially addressed in laminar
flow~\cite{odell1986flow}.  Accurate predictions in turbulent flows
require extending such work to strongly fluctuating environments and
interpreting them within a statistical framework. Remarkably, a
similar dynamics holds true for fibers immersed in highly fluctuating
environments. For instance, it was shown in
Ref.~\cite{allende2018stretching}, that the dynamics of inextensible
fibers that are smaller than the Kolmogorov scale in turbulent flow
follow most of the time that of stiff rods. Deviations occur when the
fibers experience strong-enough local compression, making them
buckle. Such events are very rare and intermittent, because of the
long-term Lagrangian correlations of turbulent velocity
gradients. During these events, the stresses experienced by the
particles can be strong enough to lead to their breakup. In fact,
turbulent flows are known to generate very large velocity gradients,
and those, in turn, may initiate a fragmentation process. Our aim here
is to provide such statistics on the mechanisms of fiber breakup in a
turbulent flow, and specifically to characterise the statistics of the
extrema of both the tension and the curvature. Those statistics
obtained with fine-scale simulations of individual fibers are used as
the basic ingredients, in order to develop accurate macroscopic
models, relevant for the above-mentioned natural and industrial
applications.  Such models predict the time evolution of the fiber
size distribution. A question that we want to address relates to the
generic nature of turbulent fragmentation processes and of the
resulting population dynamics. Turbulent fluctuations are indeed
expected to be sufficiently generic to ensure universal behaviours, as
for instance observed in~\cite{domokos2015universality} for the
fragmentation of cracking solids.

To address this problem, we resort to numerical approaches that 
couple highly-resolved turbulent flow simulations to fiber dynamics 
simulations using the slender-body equation (see below). We focus on
the case of inextensible fibers that are brittle, smaller than the 
Kolmogorov scale and that do not have an effect on the fluid.

This paper is organised as follows. In \S 2, we give a brief
description of our settings, including the slender body theory used to
model fibers and the numerical tools used to simulate their dynamics
in turbulent flow.  We moreover give an overview of the mechanisms
pertaining to fiber fragmentation (tensile failure and flexural
failure). In \S 3, we investigate tensile failure and show that it
occurs when the fiber is straight. Consequently, the tension is always
maximal at its middle and the fiber always breaks in two equal
pieces. We give also predictions on the rate at which such failures
happen and compare them to numerics. In \S 4 we turn to flexural
failure that happens when the fiber buckles. Thanks to a linear
analysis of this instability, we obtain predictions on the associated
breakup rates and on the resulting size distribution. Finally, in \S 5 we
summarise our findings and draw some perspectives. 

\section{Model and numerical method}
\label{sec:fragmentation}

The objective is to investigate fragmentation processes in a
fully-developed, homogeneous, isotropic turbulent flow.  To that aim,
we use direct numerical simulations of the three-dimensional
incompressible Navier--Stokes equation. We use the pseudo-spectral
solver \emph{LaTu} with $4096^3$ collocation points and a third-order
Runge--Kutta time marching~\cite{homann2007impact}. A force is added
at each time step to keep the kinetic energy constant in the two first
Fourier shells. This leads the velocity field to reach a statistically
stationary, homogeneous, isotropic turbulent state. The Eulerian
parameters of the simulation are summarised in Tab.~\ref{table_num}.

\begin{table}[!h]
  \begin{center}
    \begin{tabular}{|c|c|c|c|c|c|c|c|c|c|}
    \hline
    $\phantom{N^{n^n}}$\hspace{-15pt} $N^3$ & $\nu$ & $\Delta t$ & $\bar{\varepsilon}$  & $\eta$ & $\tau_\eta$ &
                                                                                                                 $u_\mathrm{rms}$
    & $L$ & $\tau_L$ & $R_\lambda $ \\ \hline
    $\phantom{N^{n^n}}$\hspace{-15pt} $4096^3$ & $10^{-5}$  &
                                                              $6\times10^{-4}$ & $3.8\times10^{-3}$ &
                                                                                                      $7.16\times10^{-4}$
                                                                                                 & $0.051$
                                                                                                               &  $0.19$
    &
      $1.86$
          & $9.68$ & $731$ \\
    \hline
    \end{tabular}
  \end{center}
  \vspace{-15pt}
  \caption{Numerical and physical parameters of the direct numerical
    simulation: $N^3$ number of collocation points, $\nu$ kinematic
    viscosity, $\Delta t$ time step, $\bar{\varepsilon}$ average kinetic
    energy dissipation rate, $\eta = \nu^{3/4}/\bar{\varepsilon}^{1/4}$
    Kolmogorov dissipative scale,
    $\tau_\eta = \nu^{1/2}/\bar{\varepsilon}^{1/2}$ Kolmogorov time,
    $u_\mathrm{rms}$ root-mean square velocity,
    $L = u_\mathrm{rms}^3/\bar{\varepsilon}$ large-eddy length scale,
    $\tau_L = L/ u_\mathrm{rms}$ large-eddy turnover time,
    $R_\lambda = \sqrt{15}\,
    u_\mathrm{rms}^2/(\nu^{1/2}\bar{\varepsilon}^{1/2})$ Taylor-based
    Reynolds number.}
  \label{table_num}
  
 \end{table}

Once in a statistical steady state, the flow is seeded with several
millions of tracers. Their dynamics is integrated with the same time
marching as the fluid and using a cubic interpolation of the velocity
field at their location.  Their trajectories, together with the fluid
velocity gradients at their location are stored with a period
$20\Delta t \approx0.23\,\tau_\eta$ for a time duration of $63000\,\Delta t
\approx 740\,\tau_\eta\approx 3.9\,\tau_L$). These data are used a
posteriori to integrate the dynamics of flexible fibers.

We consider inextensible, inertialess fibers, which are passively
transported by the flow, and hence do not influence the dynamics of
the advecting fluid. We use the model of the \emph{local} slender-body
theory describing the motion of an inextensible, inertialess
Euler--Bernoulli beam immersed in a viscous fluid (see, \textit{e.g.},
\cite{lindner2015elastic}). This model assumes that the fiber position
is described by a curve $s\mapsto \mathbf{X}(s,t)$, parametrised by
the arc-length coordinate $s\in[-\Lf/2,\Lf/2]$, such that
\begin{eqnarray}
  &&\partial_t \mathbf{X} = \mathbf{u}(\mathbf{X},t)+
     \frac{1}{\mu}\,\mathbb{D}\,
     \left[\partial_s(T\,\partial_s \mathbf{X})
     - E\,\partial_s^4\mathbf{X}\right], \quad
     \mbox{with} \quad
     |\partial_s\mathbf{X}|^2 =
     1  \label{eq:SBT} \\ 
  && \mbox{where} \quad \mu =\frac{8\pi
     \,\rho_{\mathrm{f}}\,\nu}{c}
     \quad\mbox{and}\quad \mathbb{D} =
     \mathbb{I} +
     \partial_s\mathbf{X}\,\partial_s
     \mathbf{X}^{\mathsf{T}}.
     \nonumber
\end{eqnarray}
This equation is supplemented with the free-end boundary conditions
$\partial_s^2\mathbf{X} = 0$ and $\partial_s^3\mathbf{X} = 0$ at
$s=\pm\ell/2$. In the above equations, $\rho_{\mathrm{f}}$ is the fluid mass density,
$E$ is the fiber's Young modulus and $c\gg1$ depends on the aspect
ratio. The fluid velocity field is denoted by $\mathbf{u}$ and the
tension $T(s,t)$ is the Lagrange multiplier associated to the
inextensibility constraint $|\partial_s \mathbf{X}|^2 = 1$. The
tension is intrinsically non local. The equation it solves is obtained
by requiring that $\partial_t |\partial_s\mathbf{X}|^2=0$ and reads
\begin{equation}
  \partial_s^2 T -\frac{1}{2} |\partial_s^2\mathbf{X}|^2\,T = 3\,E\,
  |\partial_s^3\mathbf{X}|^2 + \frac{7}{2}E\,
  \partial_s^2\mathbf{X}^{\mathsf{T}} \partial_s^4\mathbf{X} -
  \frac{\mu}{2}\, \partial_s\mathbf{X}^{\mathsf{T}}
  \mathbb{A}(\mathbf{X},t)\,\partial_s\mathbf{X},
  \label{eq:tension}
\end{equation}
with the boundary conditions $T=0$ at $s = \pm\ell/2$.  Here
$\mathbb{A}$ denotes the velocity gradient
$\mathbb{A}_{ij}(\mathbf{X},t) = \partial_j u_i(\mathbf{X} ,t)$. This
equation is equivalent to the Poisson equation satisfied by pressure
in incompressible fluid dynamics.

We assume that the fibers have a length $\Lf$ much smaller than the
Kolmogorov dissipative scale $\eta$. It is easily checked that their
center of mass $\Xg(t)$ then follow the dynamics of simple tracers,
namely $\mathrm{d}\Xg/\mathrm{d}t = \mathbf{u}(\Xg,t)$. Moreover, the
fluid velocity variations along the fibers can be linearised,
$\mathbf{u}(\mathbf{X},t) \approx \mathbf{u}( \Xg ,t)+\mathbb{A}(\Xg
,t)\,(\mathbf{ X}\!-\!\Xg)$ with a local velocity gradient
$\mathbb{A}$ that is constant along the fiber. Under these
assumptions, we integrate the local slender-body equation
(\ref{eq:SBT}) along the above-mentioned tracer trajectories, using
the finite-difference scheme of~\cite{tornberg2004simulating} with
$N = 201$ grid points along the fiber arc-length. The inextensibility
constraint is enforced by a penalization method. Time marching uses a
semi-implicit Adams-Bashforth scheme with a Lagrangian time step
$\Delta t_\mathrm{fib} = 2.5 \times10^{-5}$. We use a linear
interpolation in time to estimate the velocity gradient at a frequency
higher than the output from the fluid simulation. Note that the time
step required for the fibers is much smaller than that of the
fluid. Indeed, we observe that, even by using a semi-implicit scheme,
the problem remains particularly stiff when $E$ is small or,
equivalently, $\mu$ is large. As we will now see, these values of the
parameters are of particular relevance.

\begin{figure}[h!]
  \centerline{
    \includegraphics[width=.45\linewidth]{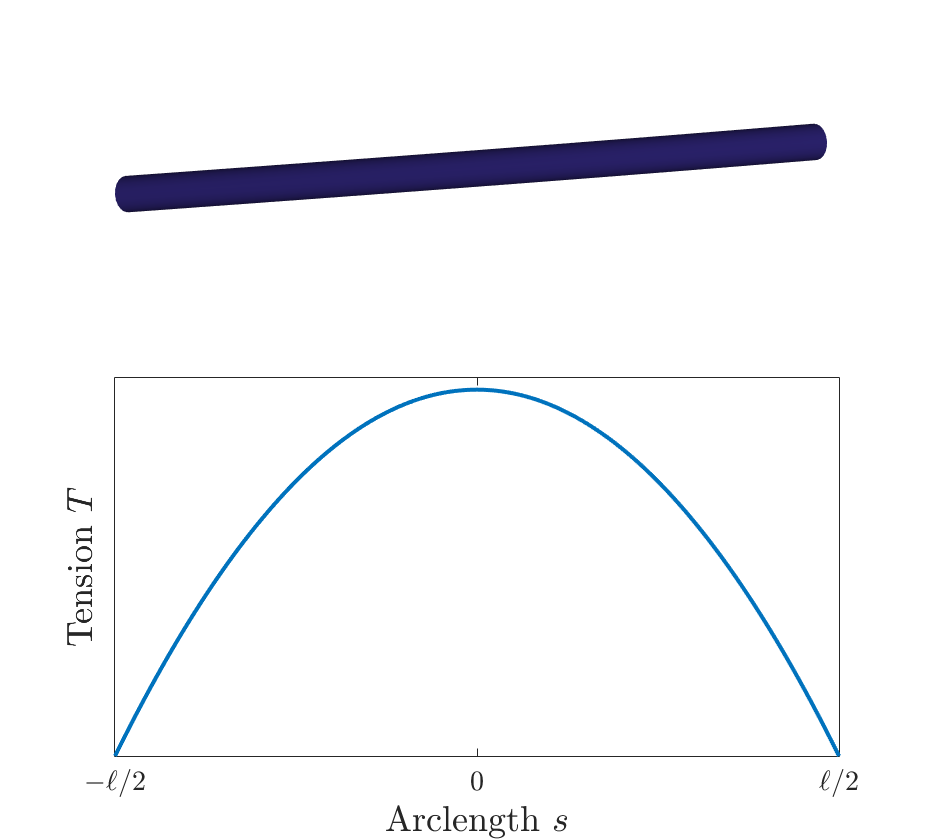}
    \qquad
    \includegraphics[width=.45\linewidth]{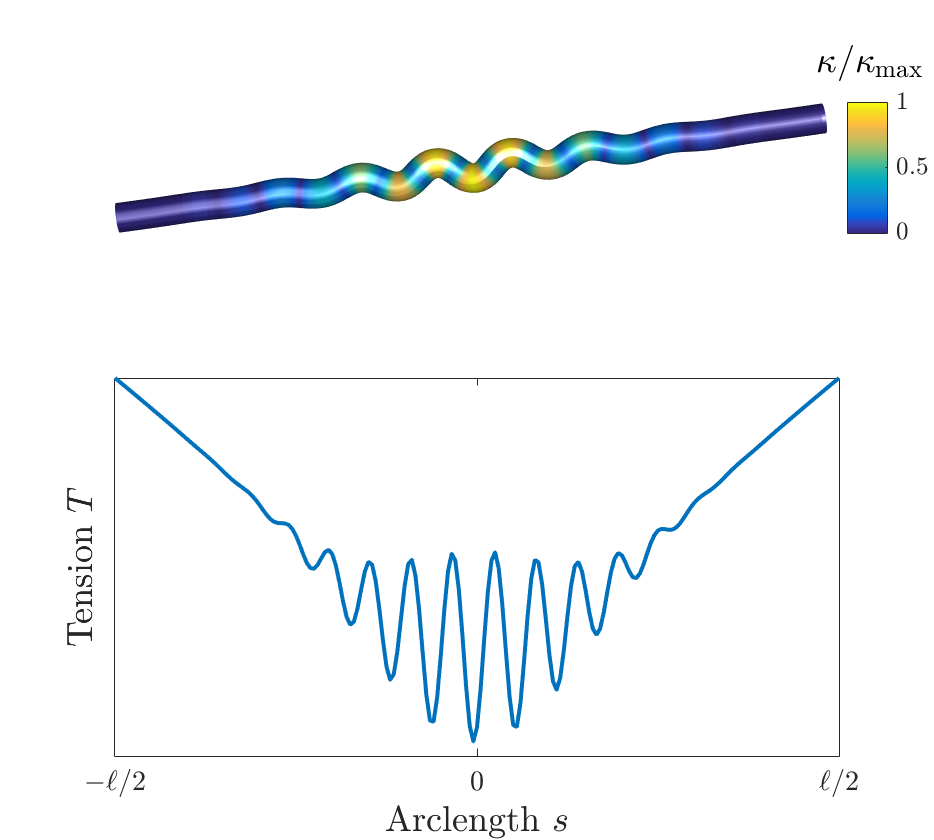}}
  \caption{\label{fig:1} Instantaneous configuration of a fiber with
    non-dimensional flexibility $\fbar=1.6\!\times\!10^5$ during a
    buckling event (right side) and as a stiff rod (left side). Top:
    the color rendering shows the fiber curvature. Bottom: tension as
    a function of the arc-length at the same instant of time}
\end{figure}

In addition to the Reynolds number $R_\lambda$ of the fluid flow, which is
prescribed very large, the dynamics of a given fiber depends on
a single dimensionless parameter only: the non-dimensional 
flexibility
\begin{equation}
  \fbar = \frac{8\pi\,\rho_{\rm f}\,\nu\,\Lf^4}{c\,E\,\tau_\eta}.
  \label{eq:defF}
\end{equation}
This parameter can be understood as the ratio between the
timescale of the fiber elastic stiffness to that of the turbulent
velocity gradients. For small values of $\fbar$, the fiber is very
rigid and behaves as a rod, while for large $\fbar$, it is very
flexible and bends. In this second case, which corresponds to long
fibers (small values of $E$ or large values of $\mu$), the dynamics
is much richer and the fiber develop into non-trivial geometrical
configurations (see top panels of Fig.~\ref{fig:1}).

When the fiber is in a fully straight state (left panel of
Fig.~\ref{fig:1}), one can assume that the tangent vector is constant,
\textit{i.e.}\/ $\partial_s\mathbf{X}(s,t) = \mathbf{p}(t)$ at all values of
$s$.  As shown in~\cite{tornberg2004simulating}, its direction
$\mathbf{p}$ is then a solution of Jeffery's equation for straight
inertialess ellipsoidal rods
\begin{equation}
  \frac{\mathrm{d}}{\mathrm{d}t}\mathbf{p} = \mathbb{A}\,\mathbf{p}
  - \dot{\gamma}\,\mathbf{p}, \quad \mbox{with}\quad \dot{\gamma}(t) =
  \mathbf{p}^{\mathsf{T}}\mathbb{A}(t)\,\mathbf{p}.
  \label{eq:jeffery}
\end{equation}
The orientation $\mathbf{p}$ is sheared and rotated by the velocity
gradient tensor $\mathbb{A}$, and the stretching/compression component
given by $\dot{\gamma}$ is removed in order to fulfill the constraint
$|\mathbf{p}|=1$. The corresponding term in the right-hand side of
(\ref{eq:jeffery}) indeed originates from the tension in the
slender-body equation (\ref{eq:SBT}). The later is obtained by from
(\ref{eq:tension}) with $\partial_s\mathbf{X} =\mathbf{p}$, leading to
\begin{equation}
  T(s,t) =  - \frac{\mu}{4}\, \dot{\gamma}(t)\left(s^2-\frac{\Lf ^2}{4}\right),
  \label{eq:tension_stretch}
\end{equation}
meaning that the tension is maximal in the middle of the fiber and
follows a parabolic shape with the arc-length coordinate $s$ (see left
panel of Fig.~\ref{fig:1}).  When the fiber is strongly compressed
($\dot{\gamma}<0$), the straight configuration might become unstable,
leading to buckling (see right panel of Fig.~\ref{fig:1}). In that
case, the tension can display several local extrema (this will be
discussed later in Section~\ref{sec:buckling_mode}).

In turbulence, fibers are most of the time in a straight state and
very rarely buckle~\cite{allende2018stretching}.  It was shown there
that the buckling instability develops when the instantaneous value of
$\dot{\gamma}$ takes large negative values. Besides, velocity
gradients in turbulent flows can experience arbitrarily large
fluctuations. This impacts the dynamics of the fiber, and in
particular the transition rates between the straight and the buckled
configurations. Explicitly, the fibers behave as stiff rods in calm
regions, and fibers bend/stretch more frequently in very fluctuating
regions.

Large turbulent fluctuations may then initiate a breakup
process. Indeed, strong stretching and compression by the flow
produces large values of the tension and curvature, respectively. As
anticipated, two mechanisms can then initiate a fragmentation:
\emph{tensile failure} when the fiber breaks because the local tension
is too high, and \emph{flexural failure} when the fiber breaks due its
curvature being too large. Large positive values of the tension
leading to tensile failure are attained when the fiber is in a fully
straight configuration and experiences a strong shear from the flow,
as shown in the left-hand side of Fig.~\ref{fig:1}. In this
configuration the bending energy is zero, thus the tension balance the
stretching due to velocity gradients. The solution of the tension is
then a parabola, where the maximum is attained in the middle of the
fiber ---\,as transpires from Eq.~(\ref{eq:tension_stretch}).

Conversely, as is illustrated on the top panel in the right side of
Fig.~\ref{fig:1}, the curvature becomes very large during buckling.
Such large values of the curvature could lead to fragmentation through
flexural failure. These instabilities are typically dominated by a
single mode with symmetric properties. Such modes depends on the
flexibility of the fiber and on the magnitude of the compression by
the flow.  Depending on which of these two mechanisms is predominant,
the fibers might break at different locations. This could imply very
different evolutions of the distribution evolution of the fiber size
distribution~\cite{grady2010length}.

\section{Fragmentation through tensile failure}
\label{sec:stretching}

We here start by investigating fragmentation due to large values of
the tension. This occurs when the maximal tension along the fiber is
larger than a critical value $T^\star$ that depends on its material
properties. To estimate the contribution of this mechanism to the
fragmentation process, we need to study both the rate
$\lambda_\mathrm{T}(T^\star)$ at which a large tension is attained
and the location where the maximum is located on the fiber.

Large tensions are reached when the fiber experiences a strong
stretching along its main axis and is thus generally in a fully
straight state. In that case, we have seen in the previous section,
\textit{c.f.}\/ Eq.~(\ref{eq:tension_stretch}), that the tension
is a concave parabola with its maximum
\begin{equation}
  T_\mathrm{max} = \frac{\mu\,\dot{\gamma} \,\Lf ^2}{16}
\end{equation}
reached at the middle of the fiber. As a consequence, a tensile
failure will always break short fibers in two equal pieces, giving a
trivial daughter distribution. It should be noted here that this is
true for fibers without molecular defects (\textit{i.e.}\/ $T^{\star}$
is constant along the fiber length).

\begin{figure}[h!]
  \centerline{
    \includegraphics[width=.6\linewidth]{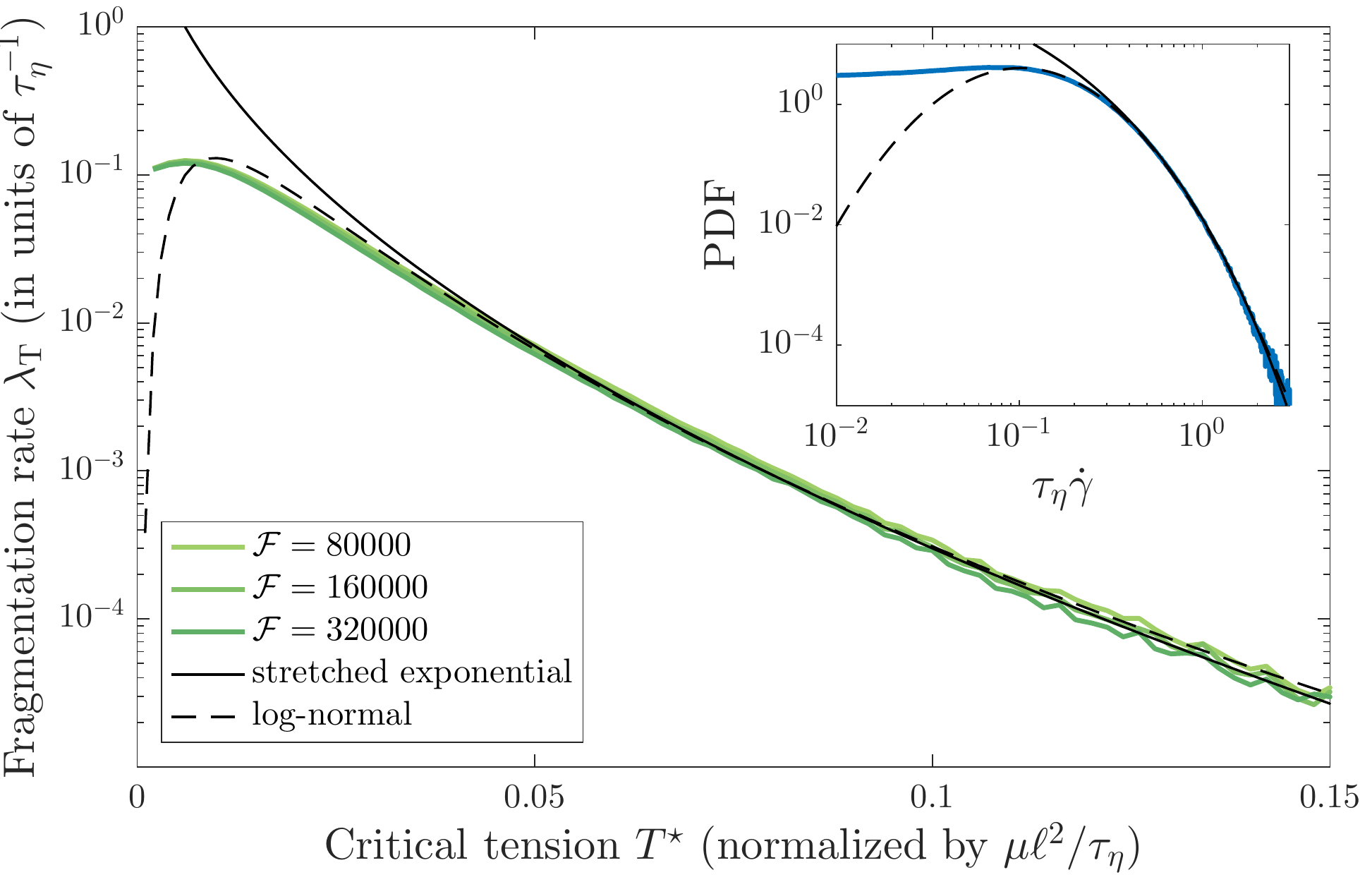}}
  \caption{Tensile fragmentation rate $\lambda_\mathrm{T}$ as a
    function of the critical tension $T^\star$ for various values of
    the non-dimensional flexibility $\fbar$. The rate has been
    non-dimensionalised by $\tau_\eta^{-1}$ while the $x$ axis
    displays $\tau_\eta\,T^\star/(\mu\ell^2)$. The solid line is the
    stretched-exponential fit (\ref{eq:Stretched_exponential}) with
    $\lambda_0 = 14\, /\tau_\eta$, $c=34$ and
    $\tau \approx \tau_\eta$. The dashed line is the log-normal fit
    (\ref{eq:lognormal}) with $\lambda_0' = 0.13/\tau_\eta$, $c'=1.12$
    and $\tau = 102\,\tau_\eta$. Inset: Probability density function
    (PDF) of the stretching rate $\dot{\gamma}$ at positive values
    (blue bold solid line), together with a stretched exponential fit
    (black thin line) and a log-normal fit (dashed line).}
  \label{fig:pdf_tension}
\end{figure}

Another consequence is that tensile failure occurs when the stretching
rate along the fiber exceeds a critical value, namely
$\dot{\gamma}>16\,T^\star/(\mu\ell^2)$, and thus when the velocity
gradient reaches strong positive values.  Turbulent gradients along
Lagrangian path are known to display very sharp fluctuations and
oscillations. This implies that the rate at which a large value of
$\dot{\gamma}$ is exceeded, is approximately proportional to the
probability distribution at this value, so that
\begin{equation}
  \lambda_\mathrm{T}(T^\star) \propto
  \mathrm{Pr}\,\left(\dot{\gamma}>16\,T^\star/(\mu\ell^2) \right).
\end{equation}
The right-hand side involves the distribution of $\dot{\gamma}$ and
thus that of fluid velocity gradients. In large-Reynolds-number
turbulence, this distribution is fairly approximated by a log-normal
with far tails which are actually closer to a stretched exponential
with an exponent $\approx 1/2$ ---\,see,
\textit{e.g.},~\cite{meneveau1991multifractal,
  kailasnath1992probability, chevillard2003lagrangian,
  donzis2008dissipation}). These two behaviours are confirmed in our
numerical simulations. The inset of Fig.~\ref{fig:pdf_tension} shows
the probability density function of the stretching rate
$\dot{\gamma}$. Its behaviour at positive values is well represented
by a log-normal distribution when $\dot{\gamma}$ is of the order of
$\tau_\eta^{-1}$ with a far tail at $\dot{\gamma}\gg\tau_\eta^{-1}$
that rather approach a stretched exponential behaviour. As stressed
in~\cite{buaria2019extreme}, the coefficients used in fits have a
non-trivial dependence upon the Reynolds number. From now on, we focus
on the case $R_\lambda = 731$, as chosen in our simulation. We will
come back to this dependence in the conclusions.

These considerations suggest to use these two possibles forms to fit
the tensile fragmentation rate $\lambda_\mathrm{T}$ as a function of
the physical parameters of the fiber. When
$ T^\star/(\mu\ell^2)\gg \tau_\eta^{-1}$, a stretched-exponential form is
expected to be more relevant, so that
\begin{equation}
  \lambda_\mathrm{T}(T^\star) \approx \lambda_0\,\exp \left[-c\,\left(
      \frac{\tau_{\eta} \,T^\star}{\mu\ell^2}\right)^{1/2}\right],
  \label{eq:Stretched_exponential}
\end{equation}
with fitting parameters a frequency, $\lambda_0$, and a dimensionless
constant $c$. As can be seen in Fig.~\ref{fig:pdf_tension}, such a
formula gives indeed a good approximation to numerical calculations of
the tensile fragmentation rates.

At $T^\star/(\mu\ell^2)\sim \tau_\eta^{-1}$, as can be seen in
Fig.~\ref{fig:pdf_tension}, the tensile fragmentation rate behaves as
a log-normal, namely
\begin{equation}
  \lambda_\mathrm{T}(T^\star) \approx \lambda_0'\,\exp \left[-c'\,\left(
      \log \frac{\tau\,T^\star}{\mu\ell^2}\right)^2\right],
  \label{eq:lognormal}
\end{equation}
where $c'$ is a dimensionless fitting constant, and $\lambda_0'$ and
$\tau$ are fitting parameters with dimensions of a frequency and a
time, respectively. The values of these parameters reported in the
caption suggest that, while $\lambda_0'$ is of the order of the
Kolmogorov timescale $\tau_\eta$, the time $\tau$ is one hundred times
larger. This can be explained by the fact that
$T_\mathrm{max}(\mu\ell^2) \sim \dot{\gamma}/16$ and in turn, as seen
in \cite{allende2018stretching}, typical values of $\dot{\gamma}$ are
of the order of $0.1/\tau_\eta$ leading to a factor of the order of
100.

To summarise this section, let us stress that tensile failure is here
entirely prescribed by the (intermittent) statistics of the velocity
gradients. This is a stylized feature of our approach, and provides a
simplified framework to study fragmentation. Note that this assumption
holds true only for small fibers.  When considering fibers larger than
the Kolmogorov length scale, the velocity gradient will not be uniform
along the fiber. This implies in particular that the fiber could be
locally stretched and compressed at the same time and breaks in pieces
of arbitrary sizes.

Also, it is important to notice that reaching large values of the
tension requires $\dot{\gamma}$ to locally exceed a critical value
that is $\propto \ell^{-2}$. Tensile failure thus becomes rarer and
rarer when fibers become smaller. In addition, the daughter
population is typically centered over the half of the mother
size. This implies that a fragmentation process involving only tensile
failure cannot efficiently lead to the fast formation of small-size
fragments. As we will see in the next section, this strongly differs 
when flexural failure is involved. In that case, many small segments 
can be created during a single event.

\section{Fragmentation through flexural failure}
\label{sec:flexural_failure}

\begin{figure}[h!]
  \begin{minipage}{.5\linewidth}
    \vspace{-210pt}
    \includegraphics[width=\linewidth]{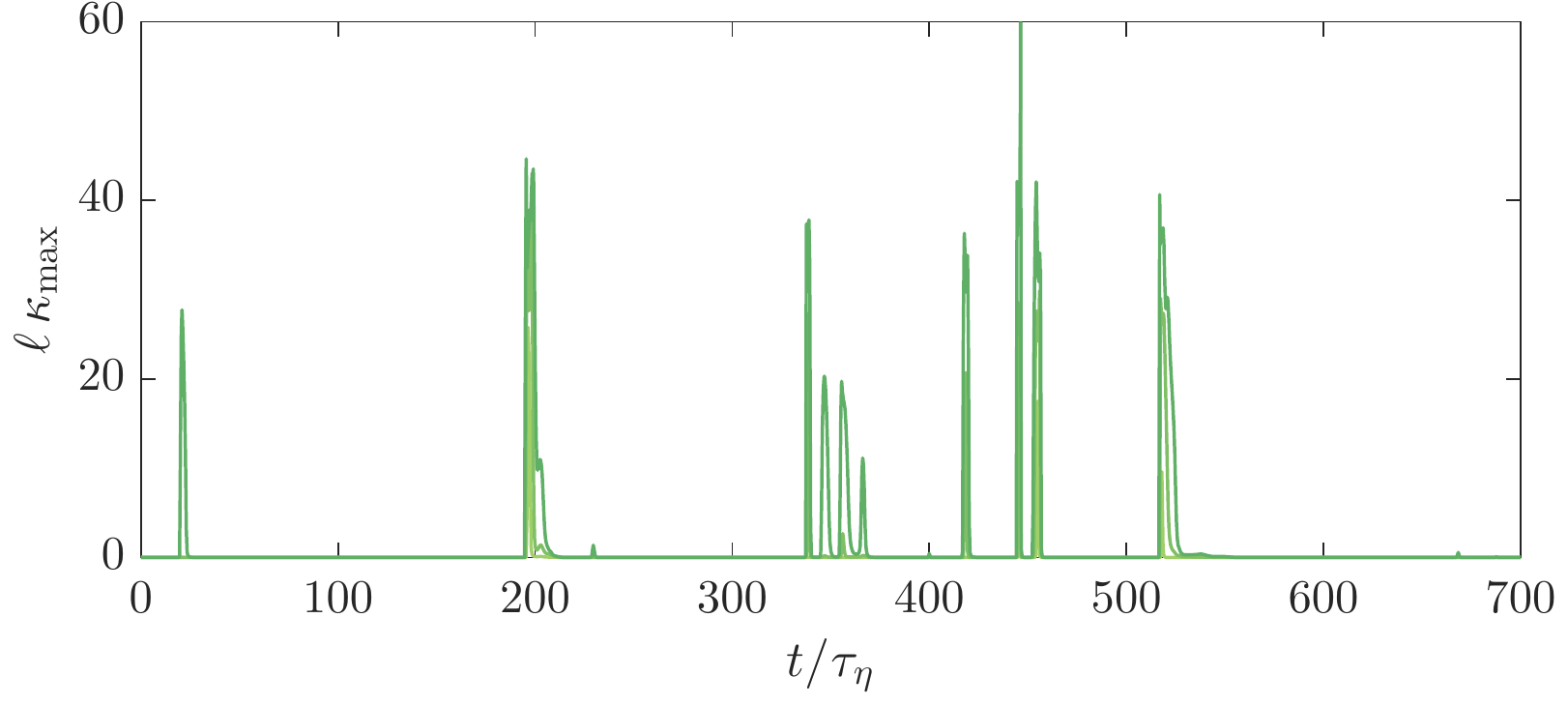}
    \includegraphics[width=\linewidth]{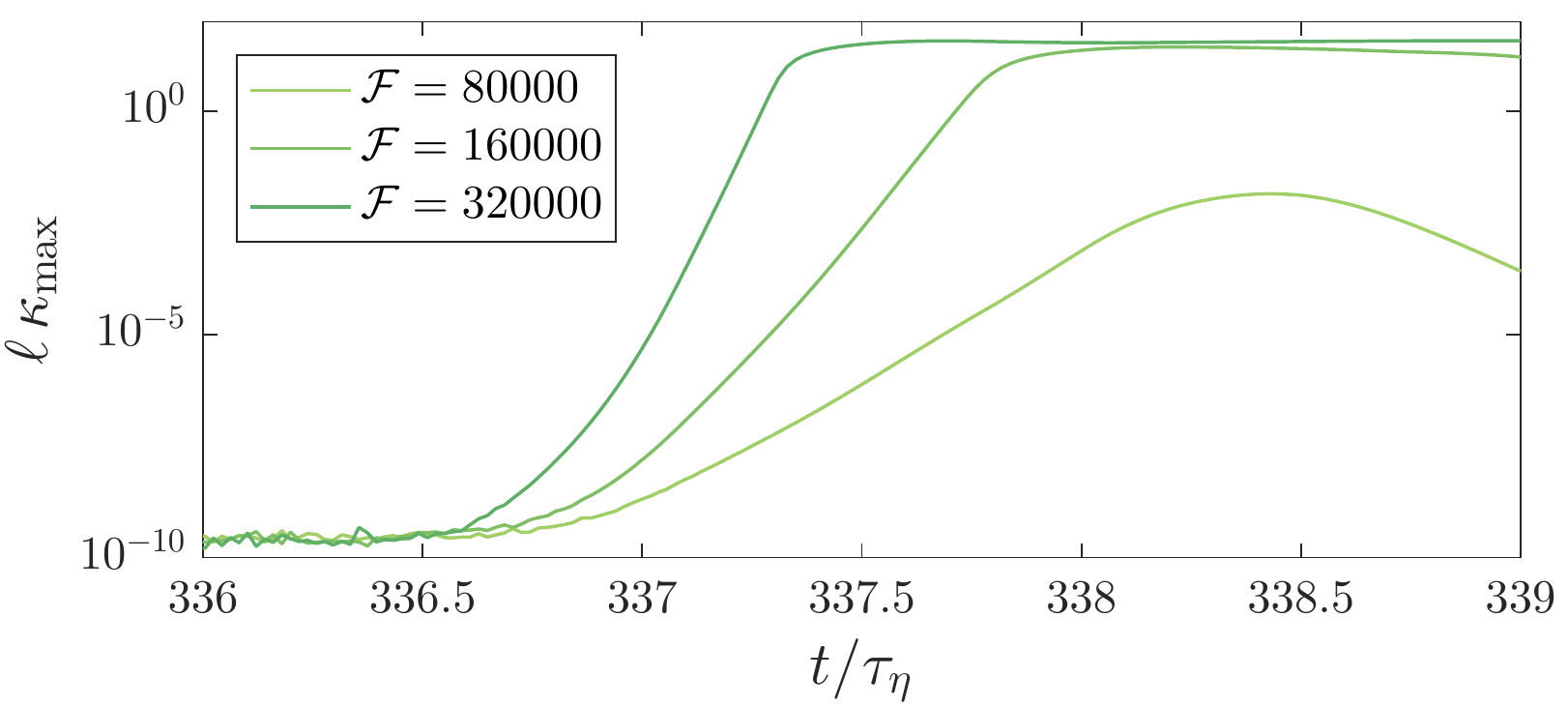}
    \end{minipage}
    \includegraphics[width=.5\linewidth]{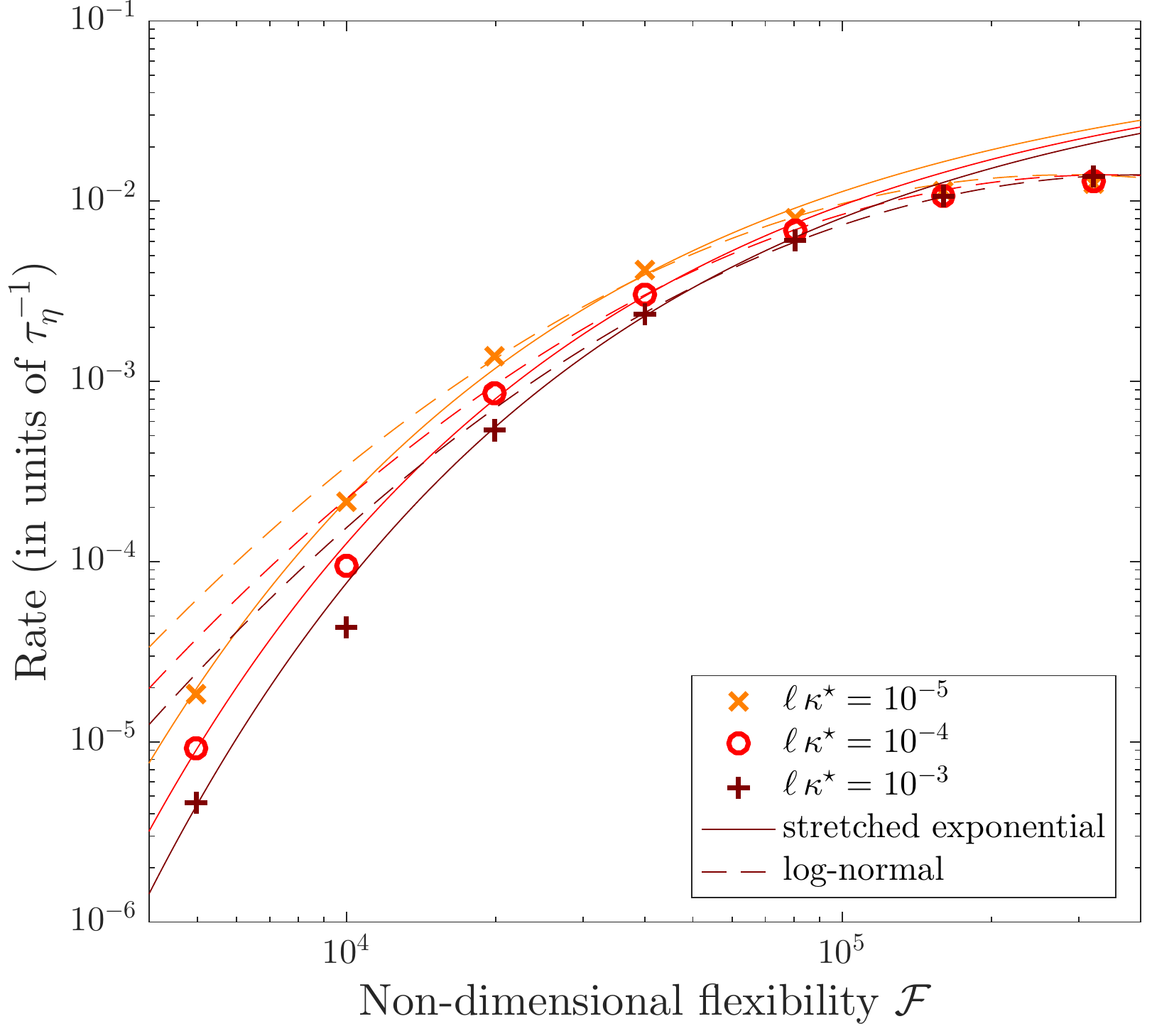}
    \caption{Left panel: Time evolution of the maximal curvature
      $\kappa_\mathrm{max}$ along a single turbulent tracer trajectory
      but for three different values of the non-dimensional
      flexibility $\mathcal{F}$. The upper part shows the full
      trajectory, while the lower is a semilogarithmic time zoom on
      the buckling event occurring at time $t\approx 337\,\tau_\eta$.
      Right panel: Rates at which the maximal curvature exceeds given
      values $\kappa^\star$ (as labelled) as a function of the
      non-dimensional flexibility. Each dataset is fitted by both a
      stretched exponential (solide curves) and by a log-normal (dashed
      curves). }
  \label{fig:energy_bend}
\end{figure}

Conversely to tensile failure, flexural failure displays a much more
complicated behaviour. Such breakups happen when the fiber is bent
and, more precisely, when the curvature becomes larger than a given
threshold. Then, there exists a location $s$ along the fiber such
that, at the breakup time $t^\star$, one has
$|\partial_s^2\mathbf{X}(s,t^\star)| \ge \kappa^\star$. Clearly, as
the curvature is continuous with respect to arc-length and time, such
a breakup occurs at the first time when the maximum of curvature
$\kappa_{\mathrm{max}} = \max_s |\partial_s^2\mathbf{X}|$ exceeds
$\kappa^\star$. As anticipated in \S\ref{sec:fragmentation}, these
events happen when the fiber undergo a buckling. Such an instability
occurs when the instantaneous value of the stretching rate
$\dot{\gamma}$ defined in Eq.~(\ref{eq:jeffery}) becomes large
negative (compression). The upper-left panel of
Fig.~\ref{fig:energy_bend} shows the time evolution of the fiber's
maximal curvature along a Lagrangian trajectory for different values
of $\fbar$. It can be seen that buckling events, for which
$\kappa_\mathrm{max}\neq 0$, are very sparse and intermittent. Such
events are separated by long periods, which can be of the order of the
large-eddy turnover time, during which the fiber is fully straight, up
to numerical precision. It is shown in
Ref.~\cite{allende2018stretching} that in turbulence, the rate at
which buckling appears is similar to an activation process. More
precisely, it was found that the fiber buckles when its instantaneous
flexibility $\fbar_{\rm loc}(t) = \tau_\eta\,|\dot{\gamma}(t)|\,\fbar$
becomes larger than a critical value $\fstar$, provided that
$\dot{\gamma}(t)<0$. This leads to estimate the buckling rate as
\begin{equation}
  \lambda_\mathrm{Buckl} \propto
  \mathrm{Pr}\,\left(\dot{\gamma}<-\fstar/ (\tau_\eta\,\fbar) \right).
  \label{eq:buckling_rate}
\end{equation}
As in the case of tensile failure rates, the distribution of the
stretching rate $\dot{\gamma}$ can be approximated either by a
log-normal or by a stretched exponential, leading to approximations of
the above formula. We expect the rate of flexural failure to be upper
bounded by this buckling rate.

However, for breakup to occur, we additionally require that the
maximum curvature exceeds $\kappa^\star$. The corresponding rates are
shown on the right-hand panel of Fig.~\ref{fig:energy_bend} as a
function of the non-dimensional flexibility $\fbar$ and for various
values of the threshold $\kappa^\star$. Log-normal and stretched
exponential functional forms give good approximations with fitting
parameters that depend on $\kappa^\star$. Understanding this
dependence requires investigating more closely the development of the
instability. The lower-left panel of Fig.~\ref{fig:energy_bend} shows
the time growth of the maximum of curvature during one of these events
for various values of the non-dimensional flexibility. One observes
that the increase is approximately exponential with a rate that
depends on $\fbar$. Small values of $\kappa^\star$ are reached during
the instability growth and it is thus needed to characterise
further this regime in order to quantify how this affect rates. The
development of the buckling instability is furthermore of importance
as flexural failure will actually not happen when the threshold is
exceeded but rather when it is for the first time. At difference with
tension that has fast fluctuations, the curvature has an on-off
behaviour. As can be seen on the lower-left panel of
Fig.~\ref{fig:energy_bend}, the growth of $\kappa_{\mathrm{max}}$ can
be followed by a period during which it remains at finite values for
quite some time. This indicates that the rates shown on the right-hand
panel of Fig.~\ref{fig:energy_bend} are actually overestimating the
actual flexural failure rates.  In the following, we provide more
accurate estimates.

\subsection{Linear analysis and relevance to turbulent flows}
\label{sec:buckling_mode}

The buckling instability occurs when the fiber, initially in a
straight configuration $\partial_s\mathbf{X} \equiv \mathbf{p}$,
experiences a strong compression by the flow. This is likely to happen
when the flow locally displays a violent shear, so that the rodlike
fiber undergoes what is known as a Jeffery
orbit~\cite{jeffery1922motion}: In that case, the rod rotates
periodically and is not aligned with the flow. During such an orbit,
the fiber alternatively experience stretching and compression along
its main axis, giving it the opportunity to buckle. Performing a
linear stability analysis for such orbits is however complex. Indeed,
upon buckling, the initially straight fiber picks a specific
trajectory among an infinite family. The
selected trajectory depends on the initial perturbation and is very
sensitive to fine sub-leading details, such as thermal
noise~\cite{munk2006dynamics}, fiber or fluid
inertia~\cite{subramanian2005inertial,einarsson2015rotation}, or, as
in our turbulent settings, the fact that the flow is not a pure shear.

As we will see in the sequel, a simplified linear stability analysis
already fairly describes buckling events, meaning that we can avoid
delving into the complicated context of Jeffery orbits. Let us
consider that the fiber experiences a time-constant compression
$\dot{\gamma} = \mathbf{p}^\mathsf{T}\mathbb{A}\,\mathbf{p}<0$ along
its direction. The base solution $\mathbf{p}(t)$ describes a rod-like
solution to the slender body equation~(\ref{eq:SBT}). We introduce a
perturbed solution as
$\mathbf{X}(s,t) = \Xg(t)+s\,\mathbf{p}(t) + \bm\chi(s,t)$, where
$\Xg(t)$ is the average position of the fiber center of mass and the
perturbation $\bm\chi$ is of small amplitude (\textit{i.e.}\/
$|\bm\chi| \ll \ell$). For buckling, we are interested in
perturbations that grow perpendicularly to the fiber direction. The
two transverse components of $\bm\chi$ are decoupled and evolve as
(see, \textit{e.g.}, \cite{lindner2015elastic})
\begin{equation}
  \frac{1}{|\dot{\gamma}|}\,\partial_t \chi =  \chi +
  s\,\partial_s \chi + \frac{1}{4}
  \left(s^2-\frac{\ell^2}{4} \right) \partial_s ^2 \chi -
  \frac{E}{\mu\,|\dot{\gamma}|}\,\partial_s ^4 \chi,
  \label{eq:linearised}
\end{equation}
with the free-end boundary conditions $\partial_s^2\chi=0$ and
$\partial_s^3\chi=0$ at $s=\pm\ell/2$. This linear equation admits
solutions of the form $\chi(s,t)=e^{ \rho\, t} \hat{\chi}(s)$, where
$\hat{\chi}$ is the eigenfunction of the right-hand side of
(\ref{eq:linearised}) associated to the eigenvalue $\rho$. Once time
is rescaled by $\dot{\gamma}^{-1}$ and arc-length by $\ell$, this
eigenvalue problem depends on a unique non-dimensional parameter
$\mathcal{F}_\mathrm{loc} = \mu\,|\dot{\gamma}|\,\ell^4/E$, which
measures the ratio between the fluid compression and the elastic 
force. Because of the presence of non-constant coefficients, there is
no straightforward way to obtain the full spectrum of eigenmodes as a
function of the dimensionless flexibility $\mathcal{F}_\mathrm{loc}$.
Still, there are two trivial solutions given by $\rho/\dot{\gamma}=1$
with $\hat{\chi} = \mathrm{const}$ and $\rho/\dot{\gamma}=2$ with
$\hat{\chi} = a\,s$ and $a$ constant. For these two unstable
modes, the fiber remains straight and does not buckle. To access more
complicated configurations, we rely on integrating numerically 
equation~(\ref{eq:linearised}).

\begin{figure}[htbp]
  \centerline{\includegraphics[width=.52\linewidth]{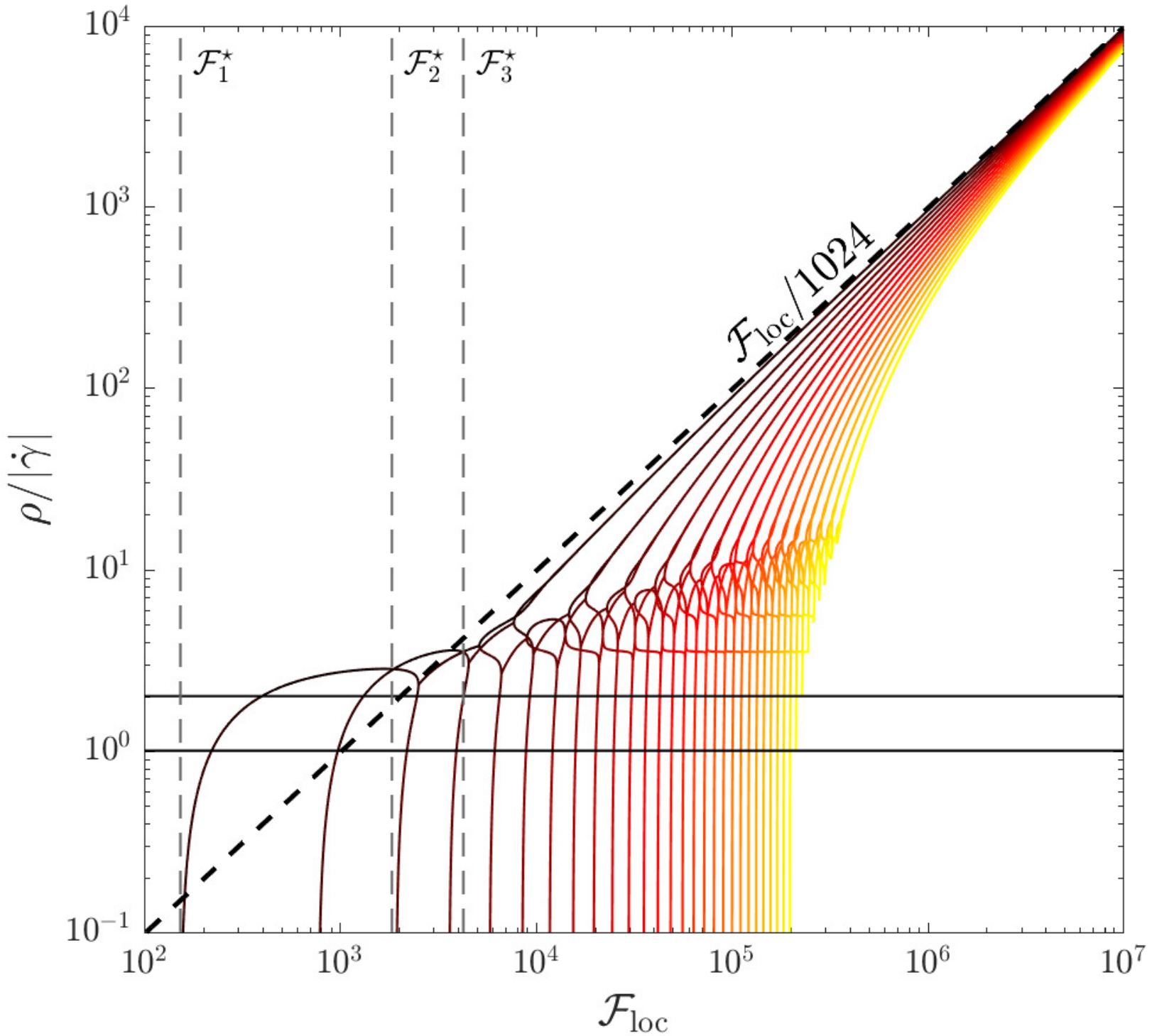}
    \hspace{-0.02\linewidth}
    \includegraphics[width=.5\linewidth]{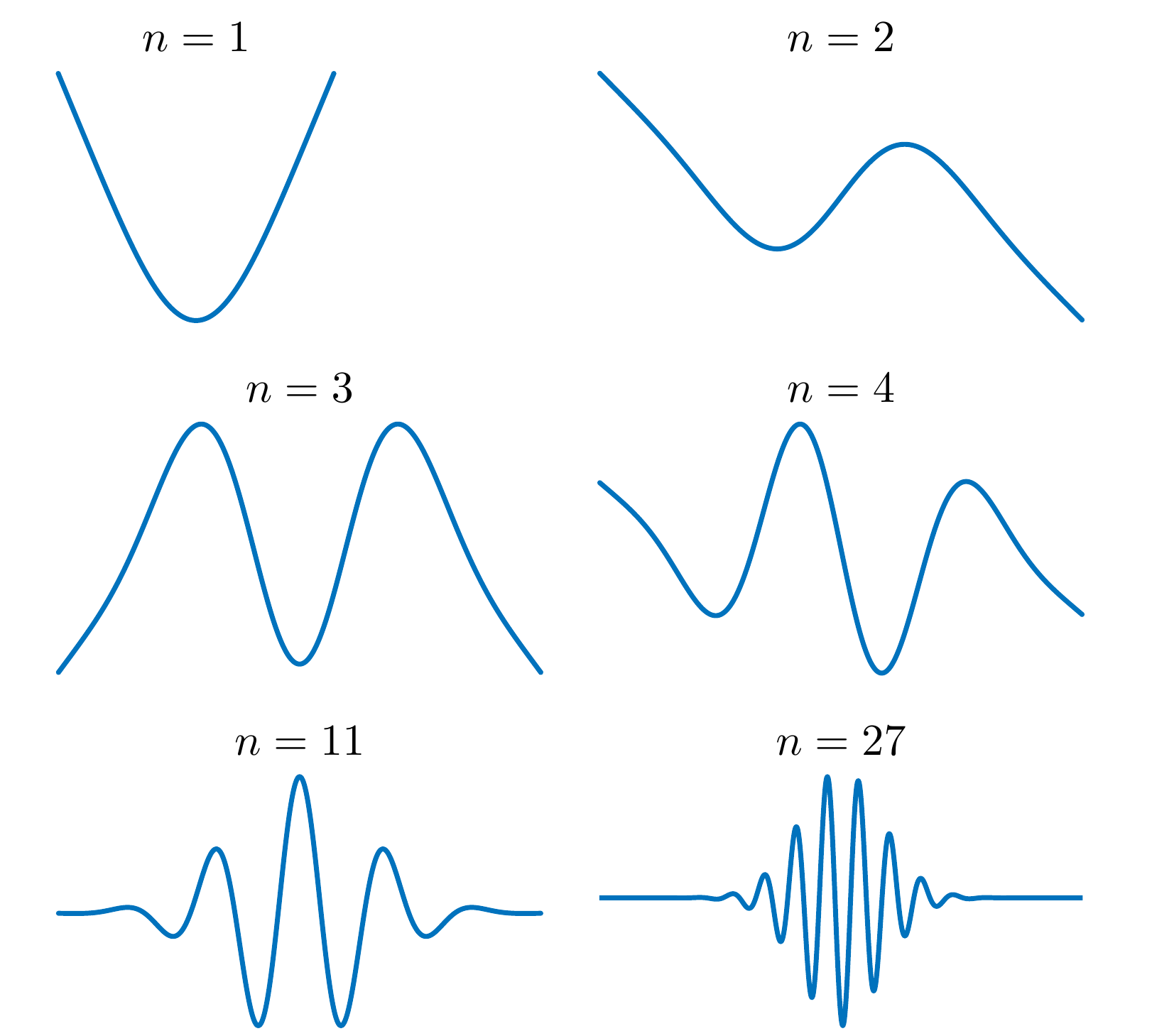}}
  \caption{Left panel: Thirty largest eigenvalues associated to the
    linear evolution (\ref{eq:linearised}) as a function of
    $\mathcal{F}_\mathrm{loc}$. The two horizontal lines at
    $\rho = \dot{\gamma}$ and $\rho = 2\dot{\gamma}$ show the two
    unstable straight modes. The three vertical lines at
    $\mathcal{F}_\mathrm{loc} =\mathcal{F}^\star_1$,
    $\mathcal{F}^\star_2$, and $\mathcal{F}^\star_3$ are bifurcations
    above which the most unstable non-straight mode is $n=1$, $2$, and
    $3$, respectively. The dashed line represents the asymptotic
    behaviour $\rho/\dot{\gamma} \simeq \mathcal{F}_\mathrm{loc}/1024$
    reached at large values. Right panel: illustrations of the fiber
    geometric state in eigenmodes of various orders $n$.}
  \label{fig:eigenvalues}
\end{figure}
The left panel of Fig.~\ref{fig:eigenvalues} represents the thirty
most unstable eigenvalues as a function of the non-dimensional
flexibility.  A first non-straight mode becomes unstable when
$\mathcal{F}_\mathrm{loc} > \mathcal{F}_1^\star\approx 153$. This
threshold is in agreement with \cite{lindner2015elastic}. This first
growing mode, labelled as $n=1$, is shown in the right-hand panel of
the figure. When $\mathcal{F}_\mathrm{loc}$ increases, there is a
sequence of bifurcations with a change of the most unstable
eigenfunction. We label these modes with the number $n$ of extrema
that $\hat{\chi}$ contains (see right-hand panel). The first
bifurcation is between order $n=1$ and $n=2$, which occurs at
$\mathcal{F}_\mathrm{loc} = \mathcal{F}_2^\star\approx 1840$. When
$\mathcal{F}_\mathrm{loc}$ further increases, the most unstable modes
are of higher order $n$. It also appears from the right panel of the
figure that the amplitude of fluctuations decreases very fast as the
arc-length coordinate $s$ gets further from the fiber center. One
finally observes that when $\mathcal{F}_\mathrm{loc}\to\infty$, the
most unstable eigenvalues grow as
$\rho/\dot{\gamma} \propto \mathcal{F}_\mathrm{loc}$.

In this asymptotics of large dimensionless flexibility, the small
parameter $\mathcal{F}_\mathrm{loc}^{-1}$ multiplies the highest-order
derivatives. This indicates that the limit is singular but could
actually be tackled using a WKB perturbative approach~(see,
\textit{e.g.}, \cite{bender-orszag}). The WKB method suggests writing
the solution as the exponential of an asymptotic series expansion
\begin{equation}
  \hat{\chi}(s) \sim \exp \left( \frac{1}{\varepsilon} \sum_{p\ge 0}
    \varepsilon^p \varphi_p (s) \right),
  \label{eq:wkb}
\end{equation}
where $\varepsilon = \mathcal{F}_\mathrm{loc}^{-\delta}$ is a small
parameter and $\varphi_p (s)$ are terms in the expansion. The 
exponent $\delta>0$ is obtained by substituting the expansion 
(\ref{eq:wkb}) in (\ref{eq:linearised}) and balancing the
leading-order terms. Far from the fiber's ends, one has
\begin{equation}
  \frac{\rho}{|\dot{\gamma}|} = \frac{1}{4\,\varepsilon^2}
  \left(s'^2-\frac{1}{4} \right) \left(\partial_{s'}
    \varphi_0\right)^2 -\frac{1}{\mathcal{F}_\mathrm{loc}
    \,\varepsilon^4}\,\left(\partial_{s'} \varphi_0\right)^4,
  \label{eq:ordre0}
\end{equation}
with $s' = s/\ell$. This gives
$\varepsilon = \mathcal{F}_\mathrm{loc}^{-1/2}$ and
$\rho/|\dot{\gamma}|\sim\mathcal{F}_\mathrm{loc}$, meaning that
$\delta = 1/2$ and confirming the observed linear behaviour of the
eigenvalues in the asymptotics of
$\mathcal{F}_\mathrm{loc}\to\infty$. Besides, when
$\mathcal{F}_\mathrm{loc}$ increases, the order $n$ of the dominant
mode becomes larger and the eigenfunction gets more localised at
$|s'|\ll 1$. This suggests expressing the dominant term as
$\varphi_0(s') = a_0 + a_1\,s'+a_2\,s'^2 + a_3\,s'^3+\cdots$. Now,
using this expansion in (\ref{eq:ordre0}) and balancing equal powers
of $s'$, one obtains to leading order
\begin{equation}
  \frac{\rho}{|\dot{\gamma}|} =
  -\frac{\mathcal{F}_\mathrm{loc}}{16}\,a_1^2-\mathcal{F}_\mathrm{loc}\,
  a_1^4,\quad\mbox{so
    that } \ a_1^2 = -\frac{1}{32}\pm\sqrt{\frac{1}{1024}-
    \frac{\rho}{|\dot{\gamma}|\,\mathcal{F}_\mathrm{loc}}}.
  \label{eq:ordre_s0}
\end{equation}
This leading term contributes an exponential behaviour
$\propto\exp(\pm\sqrt{\mathcal{F}_\mathrm{loc}}\, a_1)$ at $s'=\pm 1$
in the eigenfunction $\hat{\chi}$. When the real part of $b$ is
non-zero, this term diverges as a function of
$\mathcal{F}_\mathrm{loc}$ and this is incompatible with the imposed
free-end boundary conditions. $a_1$ should thus be a pure imaginary
number. This means that $a_1^2$ is real negative, and with
(\ref{eq:ordre_s0}), we get necessarily
$\rho/\dot{\gamma}<\mathcal{F}_\mathrm{loc}/1024$, the maximal
eigenvalue corresponding to the case when the bound is reached. This
prediction gives the value $1/1024$ for the constant of the linear
behaviour of $\rho$, in agreement with the measurements reported in
Fig.~\ref{fig:eigenvalues}. Note that for this specific eigenvalue,
one obtains $a_1=\mathrm{i}/\sqrt{32}$ ($\mathrm{i}$ being the
imaginary unit).

Moving on to higher-order terms, one can easily check that 
contributions of the order of $s'$ vanish for the above value of 
$a_1$. As to the terms $\propto s'^2$, they give
\begin{equation}
  0 = \frac{1}{4}a_1^2-\frac{1}{16}(4\,a_2^2+6\,a_1a_3) -\left(
    24\,a_1^2a_2^2+12\,a_1^3a_3\right).
\end{equation}
Using $a_1=\mathrm{i}/\sqrt{32}$, we get $a_2 = -1/8$. This finally
leads to writing
$\varphi_0\approx a_0 + \mathrm{i}\,s'/\sqrt{32} - s'^2/8$.

\begin{figure}[t]
  \centerline{\includegraphics[width=0.9\linewidth]{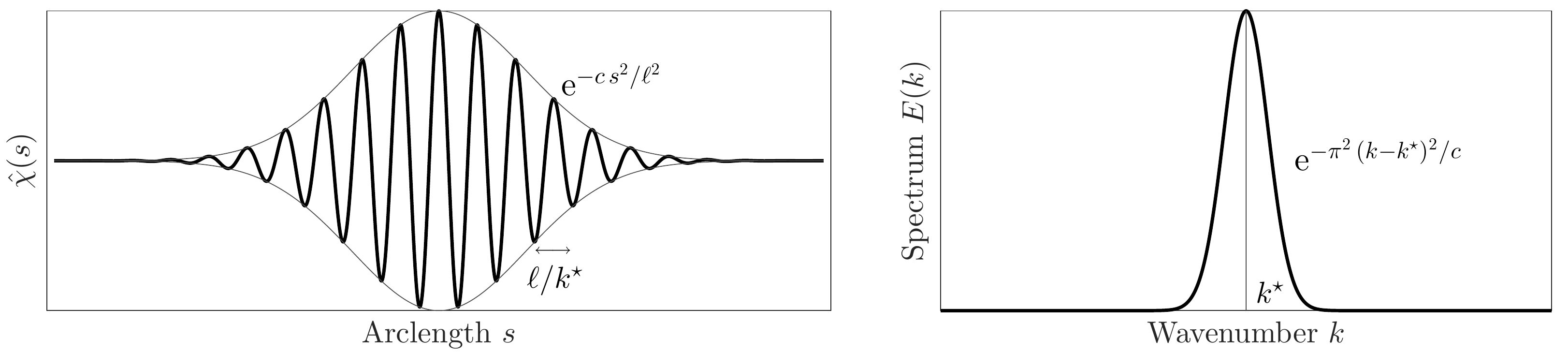}}
  \caption{Left: Typical asymptotic shape of the eigenfunctions
    consisting of a fast oscillation with a Gaussian envelope. Right:
    Spectrum $E(k) = |\mathbb{F}[\hat\chi](k)|^2$ of the same
    function, where $\mathbb{F}[\hat\chi]$ designates the Fourier
    transform of $\hat\chi$.  It is a Gaussian centered at $k=k^\star$
    with variance $c/(2\pi^2)$.}
  \label{fig:phys_Fourier}
\end{figure}

\begin{figure}[htbp]
  \centerline{
    \includegraphics[width=.51\linewidth]{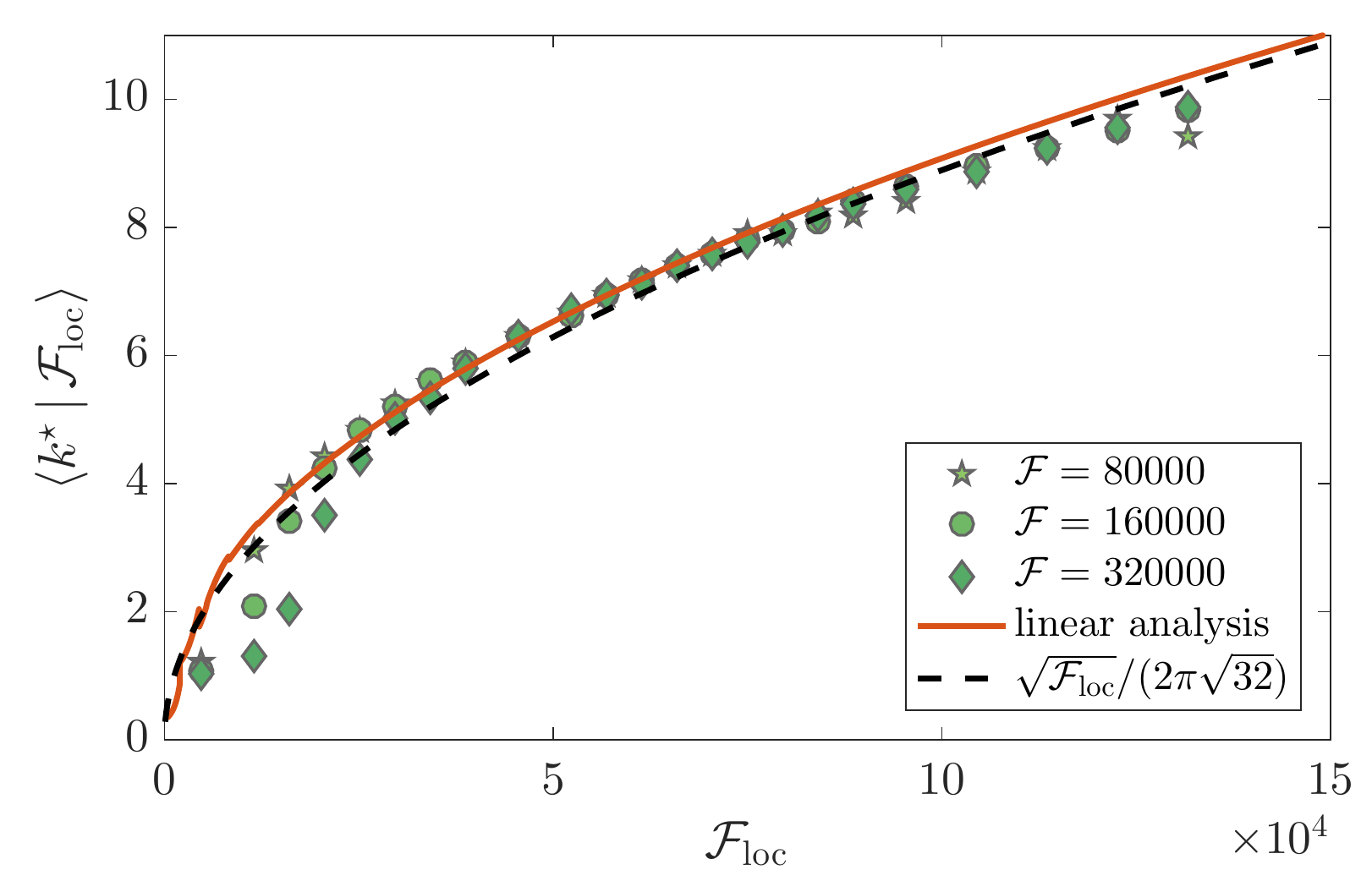}
    \hspace{-0.02\linewidth}
   \includegraphics[width=.51\linewidth]{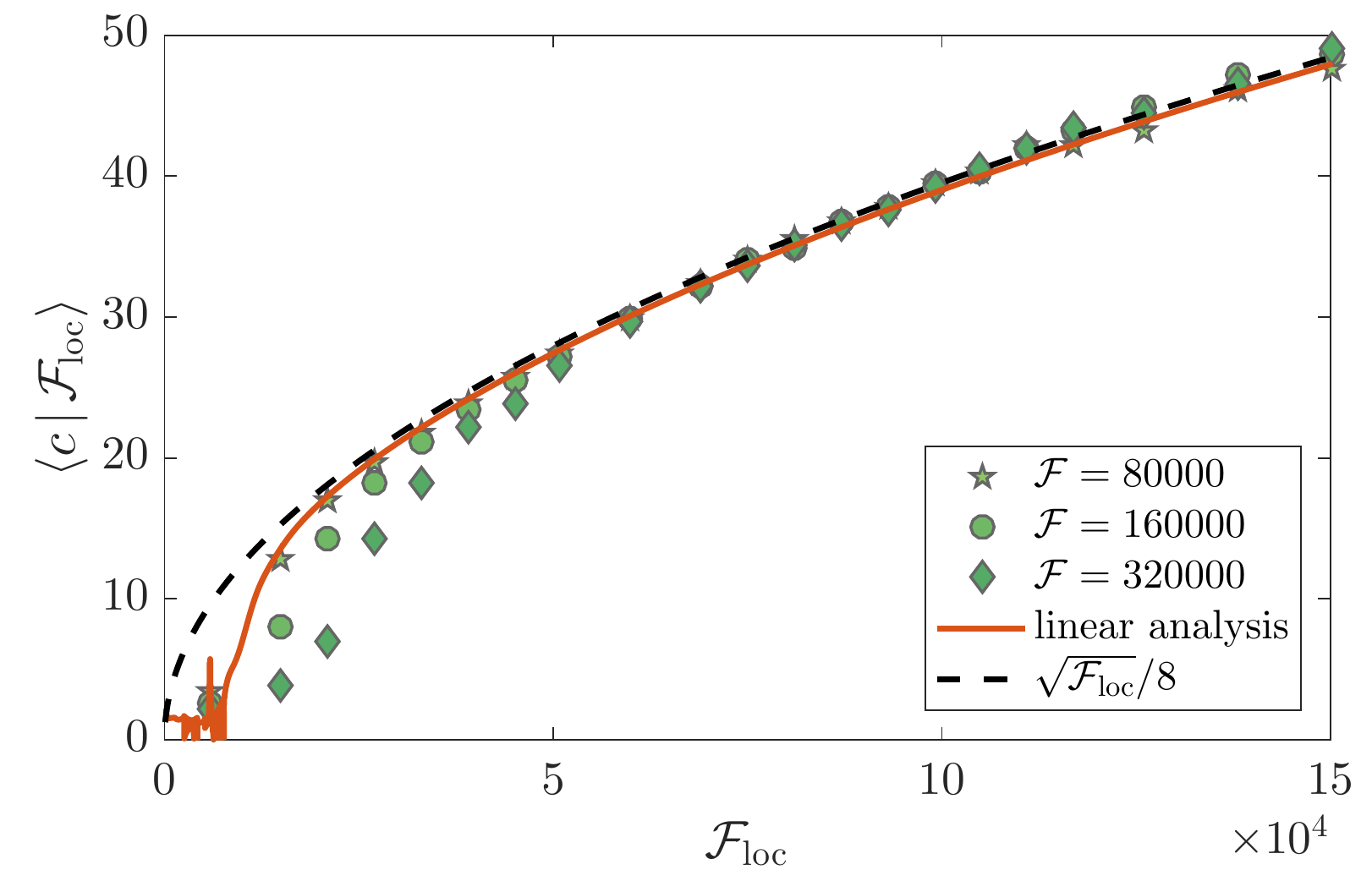}}
 \caption{Numerical estimates of the wavenumber $k^\star$ (Left) and
   of the coefficient $c$ of the Gaussian envelope (Right)
   characterising the most unstable mode. Results are shown both for
   the numerical integration of the linearised dynamics (red solid
   curves) and for turbulent data (symbols associated to various
   $\fbar$, as labelled). The asymptotic predictions
   (\ref{eq:predictions-k-c}) are shown as black dashed curves.}
  \label{fig:buckling_mode_turb}
\end{figure}

This asymptotic analysis suggest to write the eigenfunction of order
$n$ as
\begin{equation}
  \hat{\chi} (s) = \left\{ \begin{array}{ll}
                             \mathrm{e}^{-c\,(s/\ell)^2}\cos(2\pi\,k^\star\,s/\ell)
                             & \mbox{for } n = 2\,k^\star+1 \mbox{ odd}, \\
                             \mathrm{e}^{-c\,(s/\ell)^2}\sin(2\pi\,k^\star\,s/\ell)
                             & \mbox{for } n = 2\,k^\star \mbox{ even},
                           \end{array} \right.
                         \label{eq:asymp-chi}
\end{equation}
whose typical shape is represented on the left-hand side of
Fig.~\ref{fig:phys_Fourier}.  The above asymptotic analysis shows that
the most unstable mode is characterised by an oscillating function
with a wavelength $k^\star$ and a Gaussian envelope with a coefficient
$c$. These two parameters are given by:
\begin{equation}
  k^\star \simeq {\sqrt{\mathcal{F}_\mathrm{loc}}}/{(2\pi\sqrt{32})} \quad
  \mbox{and} \quad c \simeq {\sqrt{\mathcal{F}_\mathrm{loc}}}/{8}.
  \label{eq:predictions-k-c}
\end{equation}
To estimate numerically $k^\star$ and $c$, we use the Fourier spectrum
of the eigenfunction. It is defined as the squared modulus of the
coefficients of the Fourier transform of $\hat{\chi}$. As illustrated
on the right-hand side of Fig.~\ref{fig:phys_Fourier}, the spectrum of
the asymptotic form (\ref{eq:asymp-chi}) is a Gaussian function of
$k$. The wavenumber $k^\star$ is approximated as the mean associated
to this distribution, while the coefficient $c$ is deduced from its
variance.  We use this approach to measure $k^\star$ and $c$ as a
function of $\mathcal{F}_\mathrm{loc}$ for the eigenfunctions obtained
numerically from the integration of the linear system
(\ref{eq:linearised}). The results are displayed as red solid curves
on Fig.~\ref{fig:buckling_mode_turb}, together with the asymptotic
predictions~(\ref{eq:predictions-k-c}) displayed as black dashed
curves. The good agreement between these curves confirm the relevance
of the asymptotic analysis at large values of
$\mathcal{F}_\mathrm{loc}$ that we consider here.

Following this linear analysis, we then perform the same kind of 
analysis to the case of fibers that follow turbulent trajectories. 
To capture the dawn of the instability, we track fibers whose 
curvature, after having almost relaxed to zero, grows again and 
exceeds a given threshold. We use the fiber's shape at the instant of
time when the threshold is first reached as an estimate of the 
growing mode. In such fluctuating settings, we make use of the 
instantaneous non-dimensional flexibility
$\mathcal{F}_\mathrm{loc} = \tau_\eta|\dot{\gamma}|\,\mathcal{F}$. For
each event, we measure $k^\star$ and $c$ from the Fourier spectrum of
the fiber shape. We then compute their average value conditioned on 
the observed value of $\mathcal{F}_\mathrm{loc}$. The resulting 
estimates are shown as symbols on Fig.~\ref{fig:buckling_mode_turb} 
for three different values of $\mathcal{F}$. Clearly, these 
measurements show that the linear analysis reported above is able to
describe the growth of buckling modes in turbulent flows, assuming 
that the relevant parameter is given by the instantaneous value of 
the non-dimensional flexibility.

\subsection{Estimates for flexural fragmentation rates in turbulence}

We now apply the above considerations to determine both the rate at
which flexural failure occurs and the resulting daughter size
distribution upon fragmentation. We assume that the fibers are
brittle, so that they break as soon as their curvature exceeds a
critical value $\kappa^\star$, which is relatively small.  Because of
that, breakup happens while the fiber is still at the beginning of a
buckling event that can be described within the linear
approximation. We moreover assume that the instability growth is given
by eigenvalues and eigenfunctions that are properly described by the
large-flexibility asymptotics of the previous subsection.

\paragraph{Fragmentation rate} The rate at which flexural failure
occurs is the rate at which the maximal curvature along a fiber
exceeds for the first time the critical value $\kappa^\star$. A first
condition for this to happen is to have the fiber developing a
buckling: this requires
$\mathcal{F}_\mathrm{loc} =\tau_\eta\,|\dot{\gamma}|\,\mathcal{F}$
becoming larger than a critical value $\mathcal{F}^\star$ at time
$t_0$, which marks the beginning of the event.  After that, the
instability grows exponentially with a rate
$\rho \simeq |\dot{\gamma}|\,\mathcal{F}_\mathrm{loc}/1024$. The
maximal curvature also follows this growth, so that
\begin{equation}
  \kappa_\mathrm{max} (t)  \simeq \kappa_\mathrm{max}(t_0)\,
  \exp\left[\frac{\tau_\eta
      \,|\dot{\gamma}|^2\,\mathcal{F}}{1024}\,(t-t_0)\right].
  \label{eq:exp-kappa}
\end{equation}
Of course, without an initial curvature nothing would happen. It
is indeed required to have initially a small deviation to the base
state for the instability to develop. In physical situations, several
effects give such perturbations, including thermal noise, material
inhomogeneities along the fibers, a small extensibility, the fluid
flow modifications due to the fiber, and more importantly, the
sub-leading turbulent fluctuations that are neglected when we assume
that the fiber samples a space-independent fluid velocity gradient.
Such effects are clearly not in the model we use. Still, in our
simulations, the instability is triggered by a small numerical noise
that comes either from the accuracy of the method, from roundoff
errors, or from the penalty approach that is used to enforce
inextensibility. This error is visible in the lower-left panel of
Fig.~\ref{fig:energy_bend} where, before the buckling starts, one
has
$\kappa_\mathrm{max} \approx \kappa_0 \approx
2\times10^{-10}\,\ell^{-1}$.

No matter how small they are, arbitrary values of curvature are not
necessarily reached by all buckling events. For instance, during the
specific event shown in the lower-left panel of
Fig.~\ref{fig:energy_bend}, the maximal curvature barely reaches
$\kappa_\mathrm{max} = 10^{-2}\,\ell^{-1}$ in the case of the fiber 
with the smallest flexibility. The growth rate is there
too small or, equivalently, compression does not last long enough.
During this very event, the two other more flexible fibers reach much
larger curvatures and saturate at
$\kappa_\mathrm{max} \approx 30\ell^{-1}$. These distinct behaviours
originate from large differences in the instability growth rates. The
time during which the fiber is compressed is completely determined,
either by the fluid flow through the Lagrangian persistence time of
velocity gradients, or by the evolution of the base orientation
$\mathbf{p}(t)$, which for instance perform a Jeffery orbit and
tumbles. In both cases, the relevant timescale during which the fiber
is compressed is of the order of $|\dot{\gamma}|^{-1}$.

A necessary condition for the fiber to break is thus that it reaches
curvatures larger than $\kappa^\star$ on a time smaller than the
compression duration $\simeq \alpha\,|\dot{\gamma}|^{-1}$, where
$\alpha$ is an order-unity constant. Using the exponential law
(\ref{eq:exp-kappa}), one should thus have
\begin{equation}
  t-t_0 \simeq \frac{1024}{\tau_\eta\,|\dot{\gamma}|^2 \,
    \mathcal{F}}\,\log(\kappa^\star/\kappa_0) < \alpha\, |\dot{\gamma}|^{-1}.
\end{equation}
This leads to the following estimate for the rate at which flexural
failure occurs
\begin{equation}
  \lambda_\mathrm{K}(\kappa^\star) \propto \mathrm{Pr}\,\left(\dot{\gamma} < -
    \left[(1024/\alpha)\,\log(\kappa^\star/\kappa_0)\right]/
    (\tau_\eta\,\mathcal{F})\right).
\end{equation}
This formula is similar to the buckling rate of
Eq.~(\ref{eq:buckling_rate}) except that, this time, the critical
non-dimensional flexibility depends on $\kappa^\star$. Also, it
suggests that $\lambda_\mathrm{K}$ is simply a function of the
dimensionless parameter
$(1/\mathcal{F})\,\log(\kappa^\star/\kappa_0)$, that can be fitted, as
before, by either a stretched exponential or a log-normal.

\begin{figure}[htbp]
  \centerline{
    \includegraphics[width=.5\linewidth]{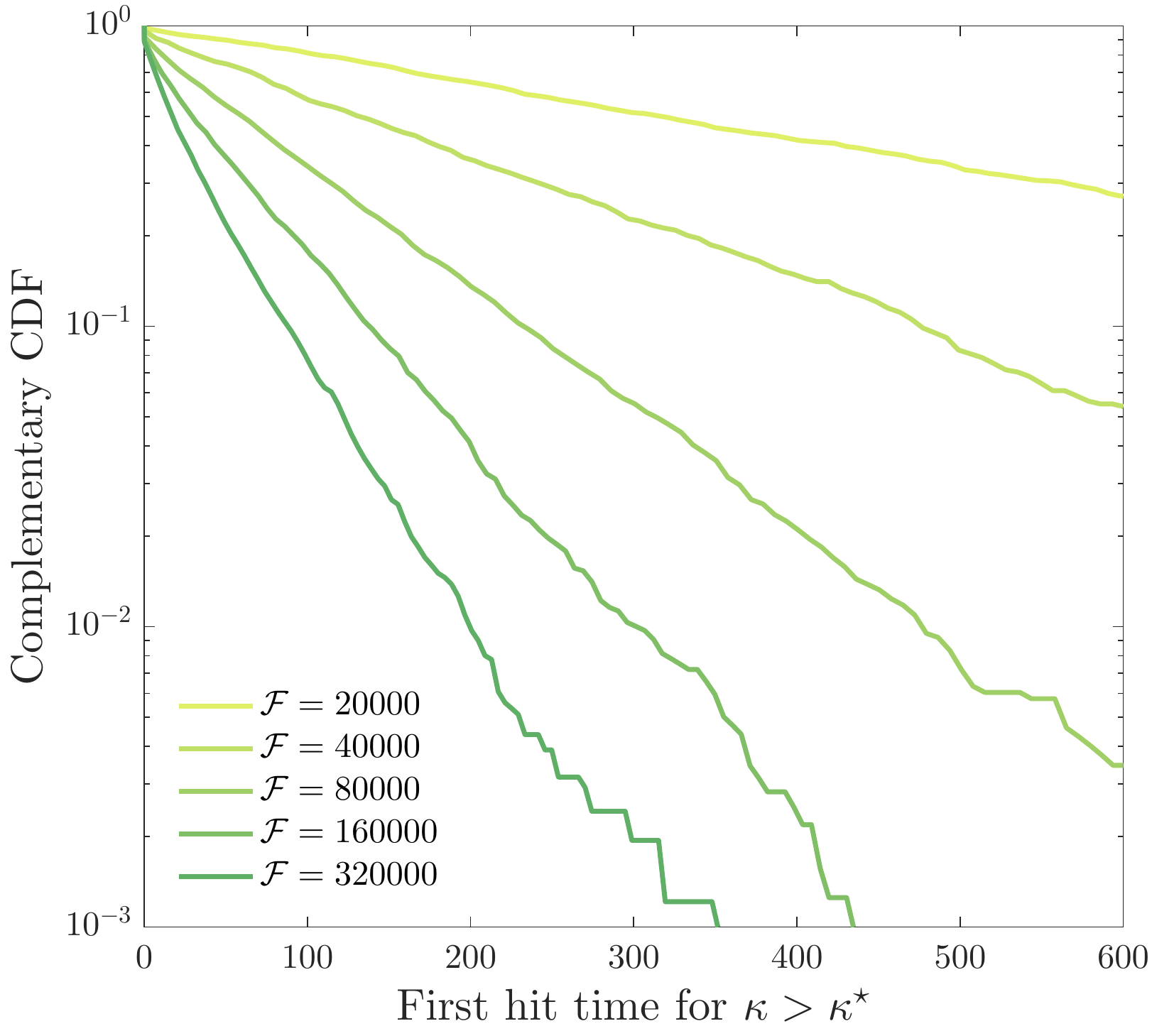}
    \includegraphics[width=.5\linewidth]{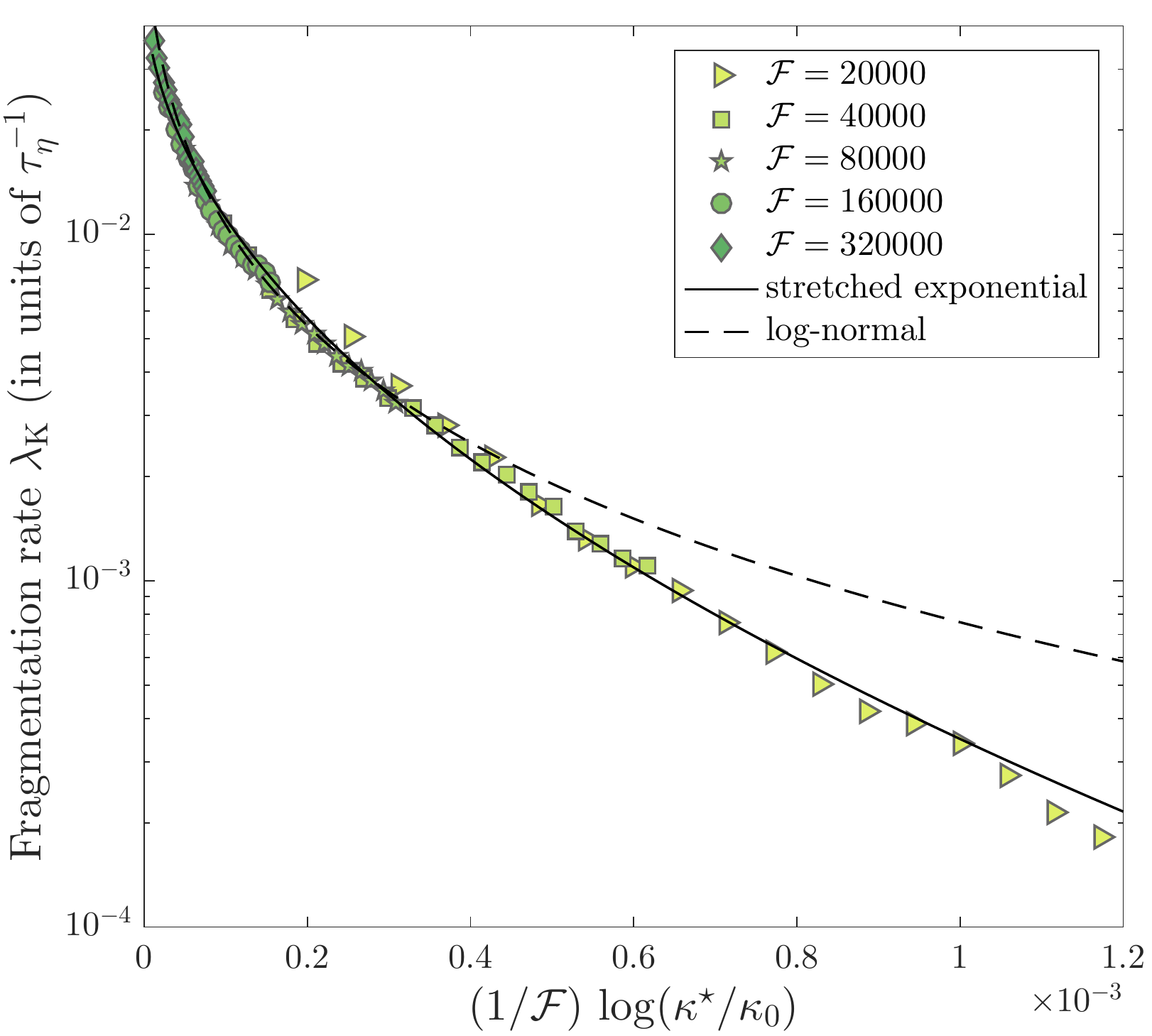}}
  \caption{Left: Complementary Cumulative Distribution Function (CDF)
    of the first time at which the fibers curvature hits a value
    $\kappa^\star = 10^{-6}$, shown for various values of the
    flexibility, as labelled. (Time is here in units of $\tau_\eta$.)
    Right: Fragmentation rate $\lambda_\mathrm{K}$ as a function of
    the dimensionless parameter
    $(1/\mathcal{F})\,\log(\kappa^\star/\kappa_0)$, with a reference
    curvature set to $\kappa_0=2\times10^{-10}$. Symbols correspond to
    different values of $\kappa^\star$ and $\mathcal{F}$. The solid
    line is the stretched-exponential fit
    (\ref{eq:Stretched_exponential_K}) with
    $\lambda_0 = 0.055\,\tau_\eta^{-1}$ and $c = 160$. The dashed line
    is the log-normal fit (\ref{eq:lognormal_K}) with
    $\lambda_0' = 0.06\, \tau_\eta^{-1}$, $c'=0.11$, and
    $a' = 13.12$.}
  \label{fig:ratecurvature}
\end{figure}
To test this prediction numerically, the main difficulty is to
estimate this rate from finite-time simulations. Depending on the
values of the fiber flexibility and of the critical curvature, the
typical time needed for the curvature to become larger than
$\kappa^\star$ can be longer than the simulation duration. Hence, to
estimate this rate, we have rather measured the probability
distribution of the first time at which $\kappa_\mathrm{max}$ hits
$\kappa^\star$. The complementary cumulative distribution function is
shown on the left panel of Fig.~\ref{fig:ratecurvature} for different
flexibilities and a fixed value of the critical curvature. Clearly,
one observes that the first-hit time follows an exponential
law. Fitting such laws gives a straightforward way to estimate
$\lambda_\mathrm{K}$ as a function of the two parameters $\mathcal{F}$
and $\kappa^\star$.
The results are shown in the right panel of
Fig.~\ref{fig:ratecurvature}. One finds that when $\mathcal{F}$ is
large-enough, all data indeed collapse when represented as a function of
the non-dimensional parameter
$(1/\mathcal{F})\,\log(\kappa^\star/\kappa_0)$.

To propose fitting formulae for this rate, we rely on approximating 
the distribution of turbulent velocity gradients by either a 
stretched exponential or a log-normal law, as was done for tensile 
failure. In the first case, we write
\begin{equation}
  \lambda_\mathrm{K}(\kappa^\star) \approx \lambda_0\,\exp \left[-c\,\left(
      \frac{\log(\kappa^\star/\kappa_0)}{\mathcal{F}}\right)^{1/2}\right],
  \label{eq:Stretched_exponential_K}
\end{equation}
with fitting parameters $\lambda_0$ (with the dimension of a
frequency) and $c$ (dimensionless). As can be seen in the left-hand
side of Fig.~\ref{fig:ratecurvature}, this a formula gives a rather
good approximation. As to the log-normal fit, it reads
\begin{equation}
  \lambda_\mathrm{K}(\kappa^\star) \approx \lambda_0'\,\exp \left[-c'\,\left[
      \log \left(\frac{\log(\kappa^\star/\kappa_0)}{\mathcal{F}}\right)+a\right]^2\right],
  \label{eq:lognormal_K}
\end{equation}
where the fitting parameters are this time a frequency $\lambda_0'$
and two dimensionless parameters $c'$ and $a'$.  As for
tension, the log-normal fit does not describe well data associated to
the tail of the distribution, that is small values of $\mathcal{F}$
or, equivalently, large values of $\kappa^\star$.  These two fits
provide estimates of the rate at which flexural failure occurs as a
function of all physical parameters, including the fiber length,
aspect ratio, Young modulus, the fluid velocity and mass density that
enter the definition of the non-dimensional
flexibility $\mathcal{F}$ given in Eq.~(\ref{eq:defF}).

\begin{figure}[h!]
  \begin{center}
    \includegraphics[width=.54\linewidth]{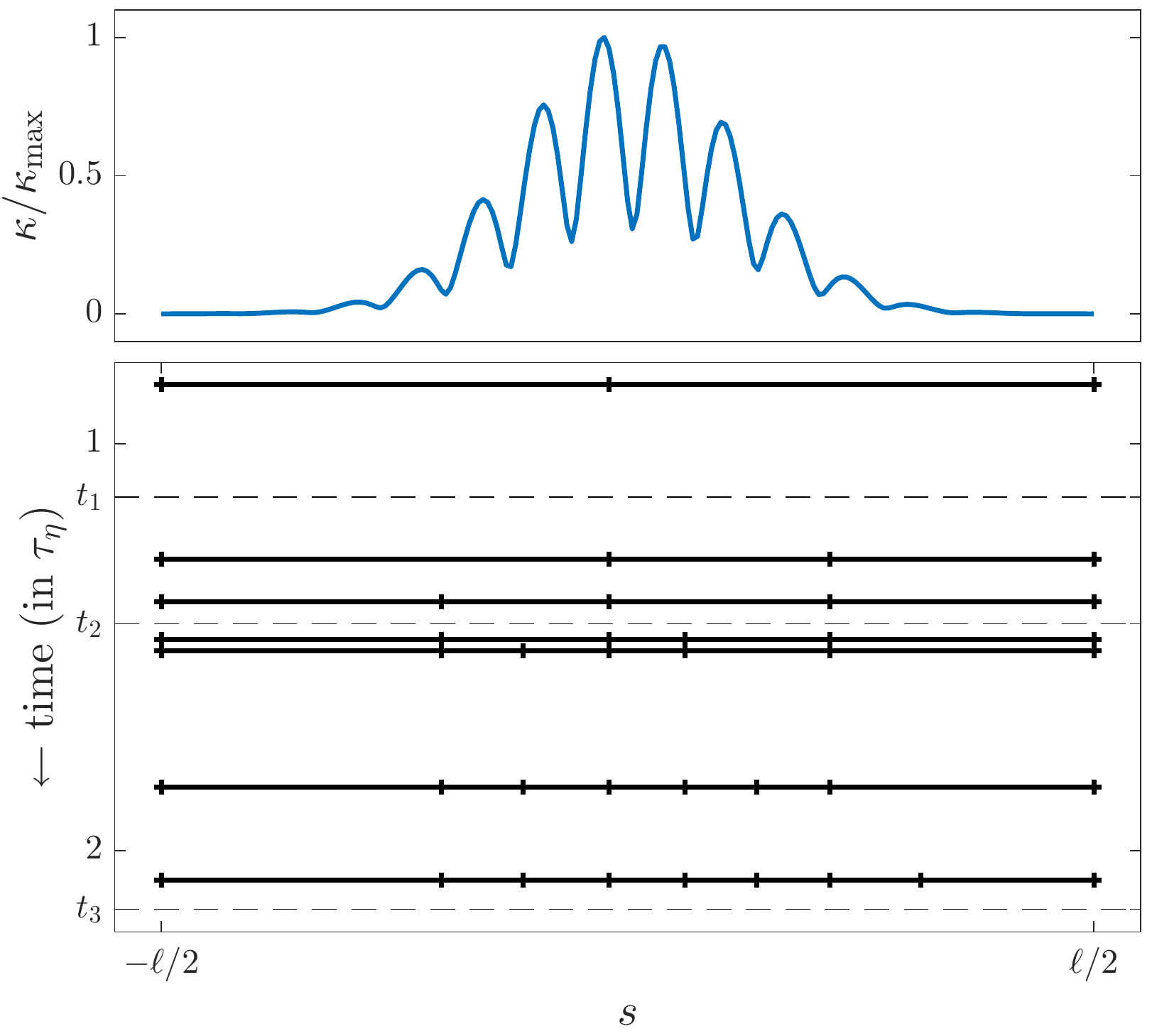}
    \begin{minipage}{.45\linewidth}
      \vspace{-250pt}
      \includegraphics[width=\linewidth]{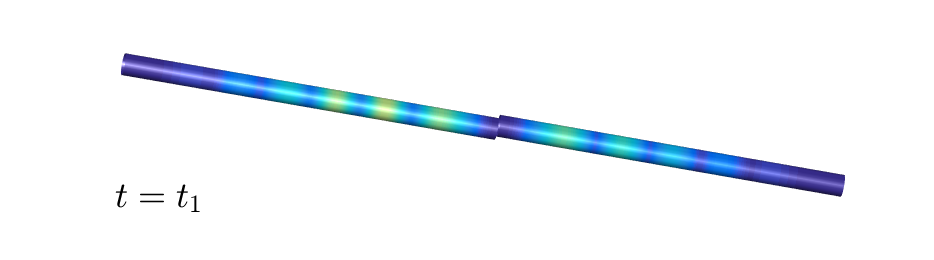}\\
      \includegraphics[width=\linewidth]{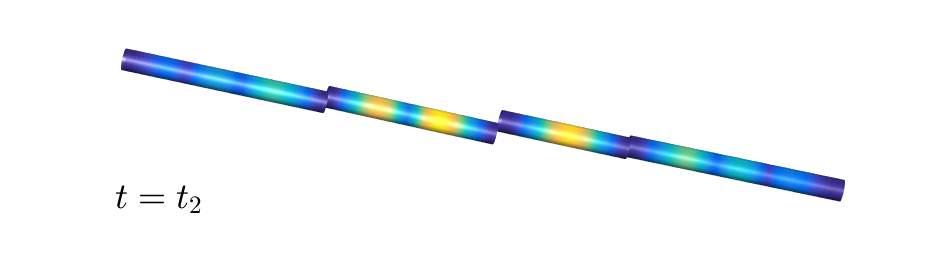}\\
      \includegraphics[width=\linewidth]{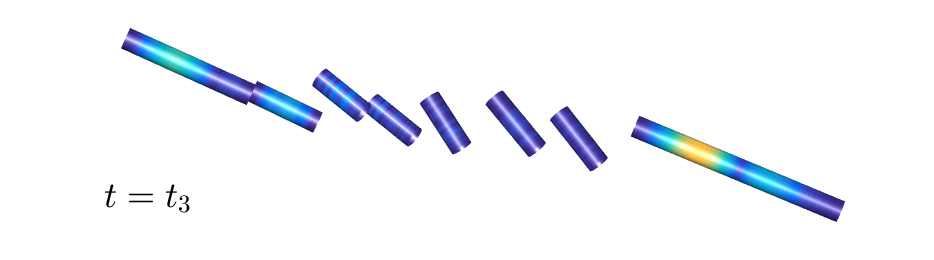}
    \end{minipage}
  \end{center}
  \vspace{-15pt}
  \caption{Evolution of a fiber during fragmentation. The top-left
    panel shows the growing mode by displaying, at the time of the
    first breakup, the fiber's curvature as a function of the
    arc-length $s$. The bottom-left panel represents the time
    evolution of the size distribution (time growing from top to
    down); Each horizontal plain line is a breakup event and segments
    correspond to fiber fragments. The right-hand side shows three
    instantanenous configurations of the fragments at time $t_1$, $t_2$
    and $t_3$. Colors code the values of curvature (from zero in dark
    blue to $\kappa^\star$ in yellow). Note that on this pseudo 3D
    representation, the arc-length $s$ runs from right to left. \label{fig:evol}}
\end{figure}

\paragraph{Daughter size distribution} We next turn our attention
to understand the resulting sizes of the fragments obtained due to
flexural failure during buckling. Up to now, by focusing on the
flexural failure rate, we have addressed only a single (the first)
breakup event.  Because the fiber is curved according to a given
buckling mode of order $n$, the location where breakup occurs clearly
depends on $n$.  When $n$ is is odd, the breakup occurs at the center
of the fiber, which breaks in two equal pieces. When $n$ is even, the
two resulting fragments have approximately sizes
$\ell\times (n/2)/(n+1)$ and $\ell\times (n/2+1)/(n+1)$. Actually,
this primary breakup is sometimes followed by successive
fragmentations. We indeed find that, because of the continuing
compression by the flow, the unstable mode keeps on bending the small
secondary pieces, so that their curvature still grows and can reach
again the critical value. This is illustrated for a specific buckling
event in Fig.~\ref{fig:evol} where we have implemented in the
numerical simulation the breakup process and the follow-up of
resulting fragments. In this case, the instability triggers the growth
of a mode of order $n=14$ (top-left panel of Fig.~\ref{fig:evol}). A
first breakup occurs at $s\approx -0.02\,\ell$, but the resulting
fragments undergo successive fragmentations. This process finally
leads to the formation of eight pieces. In this daughter distribution,
six fragments have sizes of the order of
$\ell/(n+1) \approx 0.07\,\ell$, the two remaining being associated to
the ends of the original fiber (bottom-left panel). As can be followed
on the right-hand panel, the locations where new breakups occur follow
the structures of the initial bending. Note that the full process
occurs on timescales of the order of $\tau_\eta$, confirming that this
corresponds to a single buckling event.

A single buckling event can hence lead to the creation of several
small pieces, depending on which wavenumber is excited. The selection
of the most unstable mode depends on the local value of the
non-dimensional flexibility
$\mathcal{F}_\mathrm{loc} = \tau_\eta\,|\dot{\gamma}|\,\mathcal{F}$,
which fluctuates with $\dot{\gamma}$. Following the results of
previous section, we expect for a given value of
$\mathcal{F}_\mathrm{loc}$ that the most unstable mode is of order
$n \simeq 2k^\star \simeq
\sqrt{\mathcal{F}_\mathrm{loc}}/(\pi\sqrt{32})$. In that case, the
daughter distribution will be peaked at
$\ell' = \ell/(n+1) \simeq
\pi\sqrt{32}\,\ell/\sqrt{\mathcal{F}_\mathrm{loc}}$. Assuming that
each buckling event leads to break the fiber in $(n+1)$ fragments of
equal size $\ell'$, we can draw an approximation for the daughter size
distribution. Hence, the probability that a fiber of length $\ell$
breaks in $\ell/\ell'$ fragments of size $\ell'$ reads
\begin{equation}
  \mathrm{Pr}(\ell\rightarrow \ell') \propto \frac{\ell/\ell'}{\tau_\eta\,\mathcal{F}}\,
  \ p_{\dot{\gamma}}\!\left (
    -32\,\pi^2 (\ell/\ell')^2/(\tau_\eta\,\mathcal{F}) \right ),
\end{equation}
where $p_{\dot{\gamma}}(\cdot)$ denotes the probability density function of
the stretching rate $\dot{\gamma}$.  Assuming as previously that the
later follows a stretched-exponential law, one obtains
\begin{equation}
  \mathrm{Pr}(\ell\rightarrow \ell') \propto \frac{\ell/\ell'}{\mathcal{F}}\,
  \mathrm{e}^{-c\,(\ell/\ell')/\sqrt{\mathcal{F}}}
\end{equation}
where $c$ is a positive constant.  This particularly simple form
suggests that the creation of small-length fragments follows an
activation-like distribution. In practical terms, this implies that
fragment sizes below $c\ell/\sqrt{F}$ are statistically irrelevant and
almost never observed. An equivalent form can be written for
log-normal statistics of $\dot{\gamma}$, namely
\begin{equation}
  \mathrm{Pr}(\ell\rightarrow \ell') \propto \frac{\ell/\ell'}{\mathcal{F}}\,
  \mathrm{e}^{-c'\,\left[\log\left((\ell/\ell')^2/\mathcal{F}\right)+a'\right]^2}
\end{equation}
with $a'$ and $c'$ constants. This second leads to the same
qualitative considerations as above.

\begin{figure}[b!]
    \begin{center}
  \includegraphics[width=0.65\linewidth]{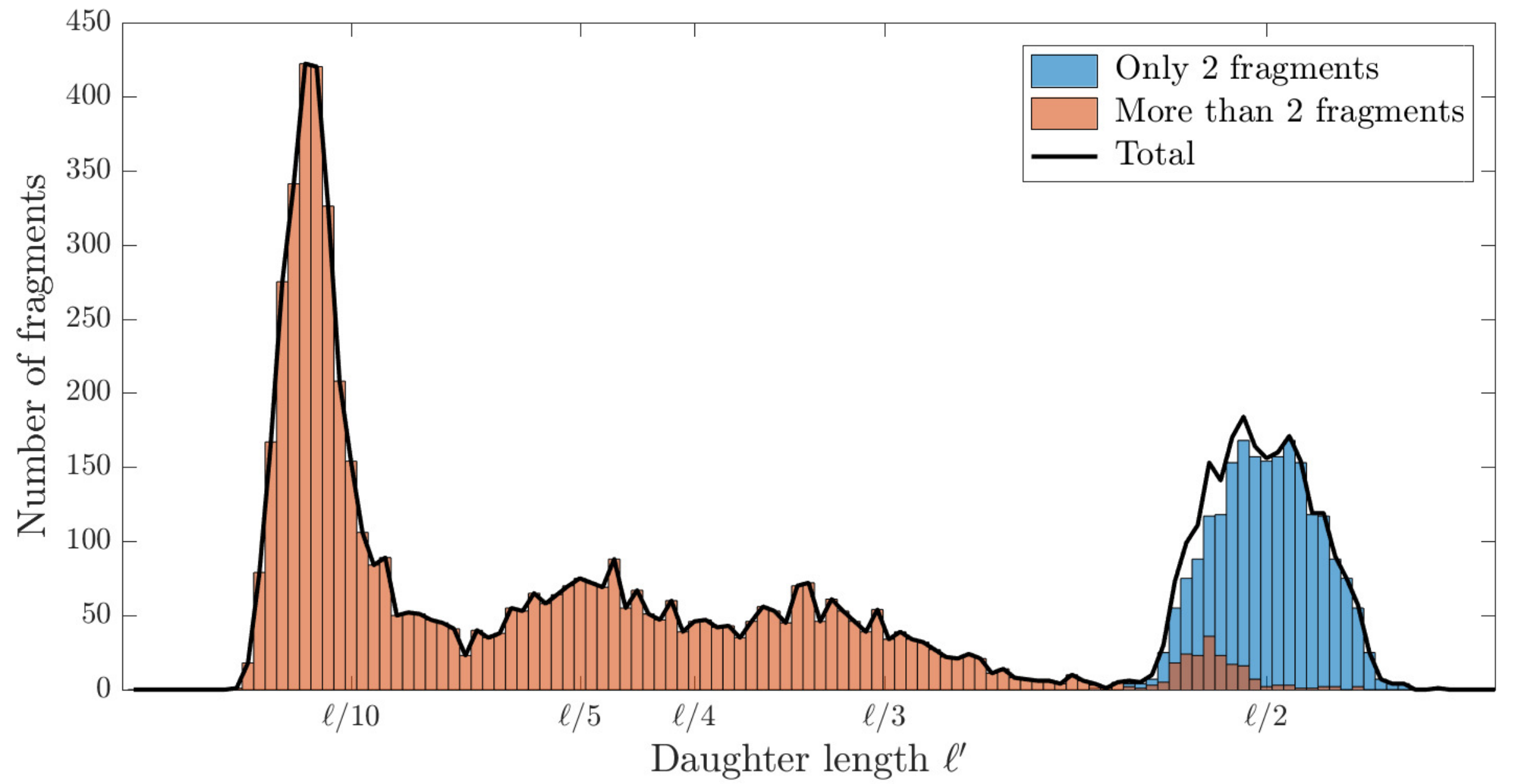}
  \caption{Histogram of the length $\ell'$ of fragments stemming from
    the first fragmentation event encountered by $2000$ fibers with
    initial length $\ell$ and a flexibility $\mathcal{F} =
    320000$.}
  \label{fig:pdf_pieces}
    \end{center}
\end{figure}
Finally, to illustrate the complexity of this process, we show in
Fig.~\ref{fig:pdf_pieces}, the fiber length distribution of fibers
after the first fragmentation series.  All fibers have initially the
same length $\ell$ and the same non-dimensional flexibility
$\mathcal{F} = 32000$. The simulation is done for a specific value of
the critical curvature ($\kappa^\star = 10^{-6}\ell^{-1}$), above
which these fibers break. A fragmentation series is defined as the set
of breakups occurring during the same buckling event, that is in a
time lag going from the initial development of the buckling
instability to the time when all fragments have relaxed to a fully
straight configuration.  The resulting distribution is clearly
multi-modal. It develops a peak at sizes $\ell'$ of the order of
$\ell/2$, corresponding to fibers that broke only once during this
series. Such events represent approximately one half of the
fragmentations. Another marked peak is present at
$\ell'\approx \ell/12$. This other maximum is an artefact of the
representation, as smaller are the segment, more numerous they
are. Clearly, the number of fragments sharply tends to zero when their
size becomes very small. Finally, the distribution displays finite
values at intermediate sizes, contributing a significative probability
of obtaining fragments of sizes $\ell/3$, $\ell/4$, $\ell/5$, etc.

\section{Conclusions and perspectives}
\label{sec:concl}

We have studied the fragmentation processes undergone by small,
inextensible, inertialess fibers in turbulent flow, focusing on
tensile and flexural failures.  In both cases, we have shown that
breakup occurs when the fiber runs into a flow region of high strain,
where it is either stretched above internal cohesive forces or
compressed, buckles and fractures under excessive bending.  By
assuming an idealised description for internal fragmentation
processes, we found that the fragmentation rates can be expressed
through the distribution of turbulent stretching rates.  Using
standard functional approximations for such probability laws, we
proposed fits for both the tensile failure and the flexural failure
rates that are calibrated and validated against the results of direct
numerical simulations in a high-Reynolds-number flow. Our analysis
emphasises the central role played by the fibers non-dimensional
flexibility in understanding how frequently fragmentation occurs.

Besides rates, we reported results on daughter size distributions upon
fragmentation. Tensile failure always occurs when the fiber is
stretched by the flow and thus has a fully straight configuration.
The tension is then maximal at its center, so that this type of
breakup always produces two fragments with equal sizes. We found that
the situation is more intricate in the case of flexural failure.
Fragmentation occurs when the fiber develops a buckling instability
and the resulting breakup process produces a size distribution that
depends on the details of the most unstable buckling mode.  By
performing a linear stability analysis, we provided estimates of this
size distribution that depend on the instantaneous fluid strain
experienced by the fiber.  This approach indicates that the number of
fragments produced upon breakup becomes larger when the fibers
experiences more violent compressions. Our analysis builds upon the
results of Vandenberghe \& Villermaux~\cite{vandenberghe2013geometry}
on how buckling affects the fragmentation of elastic slender
bodies. Specifically, they show the existence of an additional effect
in the fragmentation process: the propagation of elastic waves after a
first breakup. The released energy is able to increase further the
curvature in the newly separated parts of the fiber, possibly leading
to successive secondary breakups and to the formation of many
small-size fragments.  In our work, we unveil another, possibly
complementary mechanism where the smaller-size fragments appear as
fingerprints of the initial most-unstable growing mode.

An appealing prospective to our work concerns the Reynolds-number
dependence of fragmentation rates and processes. While our numerics
limited themselves to a single level of turbulence, our analysis can
be easily extended to encompass intermittent descriptions of turbulent
statistics.  At a qualitative level, one expects more violent
fluctuations as the Reynolds number increases.  This implies that the
rare events leading to the production of many small fragments should
have an increasing statistical relevance.  At a quantitative level,
one can for instance apply the recent work of Buaria \textit{et
  al.}~\cite{buaria2019extreme} who carefully investigated the
Reynolds-number dependence of the probability distributions of
velocity gradients.  Their findings can be straightforwardly used in
our approach to write an explicit Reynolds-number dependence of the
fragmentation rates and associated daughter size distributions.

Finally, it is important to stress again that, in this work, we have
oversimplified the microscopical breakup processes by considering that
the fibers material is brittle and that fibers are free of any
molecular defects. In most realistic settings, the threshold value for
each breakup mechanism may vary along the fiber length while plastic
effect cannot be neglected, meaning that flexural failure may occur at
locations that have been bended several times in the fiber's
history. Even if they involve an extra parametrisation, such effects
can be easily implemented and investigated numerically (see,
\textit{e.g.},~\cite{marchioli2015turbulent}).  We expect in
particular interesting impacts of the non-trivial time distribution of
violent fluctuations. The turbulent fluid strain that is experienced
by a fiber along its Lagrangian trajectory is typically very
intermittent in time, so that buckling events are strongly correlated
among each other.  Nevertheless, generalising to such settings the
analysis that has been developed here represents a real challenge. The
strongest bending is indeed expected to occur when the buckling
instability has saturated, questioning in that case the relevance of
the linear analysis.

\section*{Acknowledgements}
We acknowledge H.~Homann and C.~Siewert for their essential help with
the numerical simulations, as well as G.~Verhille and B.~Favier for
discussions.  This work was performed using HPC resources from
GENCI-TGCC (Grant t2016-2as027).  SA~has been supported by EDF R\&D
(projects PTHL of MFEE and VERONA of LNHE) and by the French
government, through the Investments for the Future project
UCA$^{\mathrm{JEDI}}$ ANR-15-IDEX-01 managed by the Agence Nationale
de la Recherche.

\end{document}